\documentclass[modern]{aastex61}

\newcommand{\RQ}{\textnormal{\scriptsize \textsc{rq}}}
\newcommand{\RL}{\textnormal{\scriptsize \textsc{rl}}}
\newcommand{\LC}{\textnormal{\scriptsize \textsc{lc}}}
\newcommand{\SUR}{\textnormal{\scriptsize \textsc{s}}}
\newcommand{\CUT}{\textnormal{\scriptsize \textsc{c}}}
\newcommand{\X}{\textnormal{\scriptsize \textsc{x}}}
\newcommand{\POLAR}{\textnormal{\scriptsize \textsc{p}}}
\newcommand{\NS}{\textnormal{\scriptsize \textsc{ns}}}
\usepackage{amsmath}

\received{mm dd, yyyy}
\revised{mm dd, yyyy}
\accepted{mm dd, yyyy}
\submitjournal{ApJ}

\shorttitle{Bayesian Radio-quietness of $\gamma$-ray Pulsars}
\shortauthors{Yu et al.}

\begin{document}

\title{Bayesian Inference on the Radio-quietness of Gamma-ray Pulsars}

\correspondingauthor{Hoi-Fung Yu}
\email{dy@kth.se}

\author[0000-0001-5643-7445]{Hoi-Fung Yu}
\affiliation{Department of Physics, KTH Royal Institute of Technology, 10691 Stockholm, Sweden}
\affiliation{Oskar Klein Centre for Cosmoparticle Physics, 10691 Stockholm, Sweden}

\author[0000-0003-1753-1660]{Chung Yue Hui}
\affiliation{Department of Astronomy and Space Science, Chungnam National University, Daejeon 34134, Korea}

\author{Albert K. H. Kong}
\affiliation{Institute of Astronomy, National Tsing Hua University, Hsinchu, Taiwan}

\author{Jumpei Takata}
\affiliation{Institute of Particle Physics and Astronomy, Huazhong University of Science and Technology, China}

\begin{abstract}

We demonstrate for the first time using a robust Bayesian 
approach to analyse the populations of radio-quiet (RQ) and 
radio-loud (RL) gamma-ray pulsars. We quantify their 
differences and obtain their distributions of the radio-cone 
opening half-angle $\delta$ and the magnetic inclination 
angle $\alpha$ by Bayesian inference. In contrast to the 
conventional frequentist point estimations that might be 
non-representative when the distribution is highly skewed 
or multi-modal, which is often the case when data points 
are scarce, Bayesian statistics displays the complete 
posterior distribution that the uncertainties can be 
readily obtained regardless of the skewness and modality. 
We found that the spin period, the magnetic field strength 
at the light cylinder, the spin-down power, the 
gamma-ray-to-X-ray flux ratio, and the spectral curvature 
significance of the two groups of pulsars exhibit 
significant differences at the 99\% level. Using Bayesian 
inference, we are able to infer the values and uncertainties 
of $\delta$ and $\alpha$ from the distribution of RQ and 
RL pulsars. We found that $\delta$ is between $10^\circ$ 
and $35^\circ$ and the distribution of $\alpha$ is skewed 
towards large values.

\end{abstract}


\section{Introduction} \label{sec:intro}

The numbers of radio-quiet (RQ) and radio-loud (RL) gamma-ray 
pulsars detected have increased rapidly during recent years 
thanks to the launch of the \textit{Fermi} Gamma-ray Space 
Telescope. The observed differences between RQ and RL pulsars 
are expected to be not intrinsic, but due to geometrical 
effects. Theoretically, the radio emissions are originated 
from the polar cap region, which is aligned with the magnetic 
axis of the pulsar \citep{Ruderman75}. On the other hand, the 
gamma-rays are generally accepted to be originated from the 
outer gap \citep[e.g.,][]{Cheng98,Takata06,Takata08}, which 
subtends a larger solid angle than the radio cones. Therefore 
the probability of the line-of-sight (LOS) to intersect with 
the gamma-ray emission region is higher. If the projection of 
the radio cone does not swap through the LOS, then the pulsar 
is seen as RQ. Thus, the magnetic inclination angle, $\alpha$, 
and the radio-cone opening half-angle, $\delta$, play 
important roles in the modeling of gamma-ray pulsars. 

Recently, \citet[][hereafter H17]{Hui17} and 
\citet[][hereafter S16]{Sokolova16} studied the observed 
and derived properties of RQ and RL gamma-ray pulsars. 
In our current study, we adopt the pulsar populations 
used in H17, which is extracted from the 117 gamma-ray 
emitting pulsars detected in The Second \textit{Fermi} 
Large Area Telescope \citep[LAT,][]{Atwood09} Catalog of 
Gamma-ray Pulsars \citep[2PC,][]{Abdo13} and The 
\textit{Fermi} LAT Third Source Catalog 
\citep[3FGL,][]{Acero15}. Nevertheless, we also examine 
the reported discrepancy of the inferred fraction of RQ 
pulsars between these two studies.

\citet{Hui17} performed a non-parametric analysis on the 
RQ and RL pulsar populations. Specifically, they applied 
the Anderson-Darling Test (the so-called A-D Test) to the 
unbinned data. They found that the magnetic field strength 
at the light cylinder $B_\LC$, the gamma-ray-to-X-ray flux 
ratio $F_\gamma/F_\X$, and the spectral curvature 
significance (CS) exhibit $>3\sigma$ difference. In the 
current paper, we try to verify or disprove their 
non-parametric frequentistic results using the advanced 
Bayesian formalism.

We apply a robust\footnote{The analysis is said to be robust 
if it is insensitive to outliers.} Bayesian statistical technique \citep{Kruschke13}
to the data in order to identify the key parameters that 
separate the groups of RQ and RL pulsars. We further 
obtain the posterior probability distribution of the 
fraction of RQ pulsars and the $\alpha$-$\delta$ 
posterior distribution using Bayesian inference. 
Comparing to conventional frequentist statistics, one 
of the advantages of Bayesian statistics is its 
ability to display the complete posterior probability 
distribution. The posterior provides information about 
the probability density, the uncertainty, modality, and 
correlation between parameters. Therefore, the 
uncertainties of the physical and statistical parameters 
can be readily obtained without the need to rely on 
complex error propagation, theoretical or numerical 
approximation, and sometimes point estimations that 
may not be representative if the distribution is highly 
skewed or multi-modal. It is also conceptually 
straightforward to integrate over the contributions 
of all but one parameter to get the marginal distribution, 
which displays complete information for the parameter 
in interest when the uncertainties of all other 
parameters are taken into account.

We detail the Bayesian analysis methods and results 
in respective subsections of Sect.~\ref{sec:bayes}. 
The conclusions and discussions are presented in 
Sect.~\ref{sec:disc}, where the physical implications 
of the results are also discussed.

\section{Bayesian Analysis Methods and Results} \label{sec:bayes}

\subsection{Difference between radio-quiet and radio-loud gamma-ray pulsars} \label{subsec:best}

\begin{deluxetable*}{ccccccc}
\tablenum{1}
\tablecaption{Bayesian Posterior Statistical Parameters\label{tab:posterior}}
\tablewidth{0pt}
\tablehead{
\colhead{Physical Parameter} & \colhead{$\mu^\RQ$} & \colhead{$\mu^\RL$} & \colhead{$\sigma^\RQ$} & 
\colhead{$\sigma^\RL$} & \colhead{$\nu$} & \colhead{$\mu^\RQ-\mu^\RL$}
}
\startdata
$\log[p$ (ms)] & $ 2.227 ^{+ 0.103 }_{- 0.101 }$ & $ 2.074 ^{+ 0.108 }_{- 0.107 }$ & $ 0.219 ^{+ 0.087 }_{- 0.066 }$ & 
$ 0.257 ^{+ 0.093 }_{- 0.074 }$ & $ 40.031 ^{+ 107.513 }_{- 37.387 }$ & $ 0.153 ^{+ 0.146 }_{- 0.148 }$ \\
$\log[\dot{p}$ (s s$^{-1}$)] & $ -13.58 ^{+ 0.33 }_{- 0.33 }$ & $ -13.33 ^{+ 0.27 }_{- 0.28 }$ & $ 0.70 ^{+ 0.29 }_{- 0.23 }$ & 
$ 0.66 ^{+ 0.25 }_{- 0.22 }$ & $ 32.42 ^{+ 102.61 }_{- 30.59 }$ & $ -0.25 ^{+ 0.44 }_{- 0.42 }$ \\
$\log[B_\LC$ (G)] & $ 3.606 ^{+ 0.320 }_{- 0.314 }$ & $ 4.118 ^{+ 0.289 }_{- 0.288 }$ & $ 0.686 ^{+ 0.269 }_{- 0.204 }$ & 
$ 0.686 ^{+ 0.243 }_{- 0.200 }$ & $ 41.088 ^{+ 106.775 }_{- 38.344 }$ & $ -0.511 ^{+ 0.423 }_{- 0.428 }$ \\
$\log[B_\SUR$ (G)] & $ 12.325 ^{+ 0.162 }_{- 0.163 }$ & $ 12.370 ^{+ 0.153 }_{- 0.151 }$ & $ 0.341 ^{+ 0.152 }_{- 0.123 }$ & 
$ 0.346 ^{+ 0.144 }_{- 0.129 }$ & $ 22.413 ^{+ 91.484 }_{- 20.764 }$ & $ -0.045 ^{+ 0.220 }_{- 0.216 }$ \\
$\log[\dot{E}$ (erg s$^{-1}$)] & $ 35.335 ^{+ 0.488 }_{- 0.473 }$ & $ 36.056 ^{+ 0.399 }_{- 0.403 }$ & $ 1.027 ^{+ 0.395 }_{- 0.308 }$ & 
$ 0.960 ^{+ 0.344 }_{- 0.269 }$ & $ 42.299 ^{+ 108.965 }_{- 39.574 }$ & $ -0.721 ^{+ 0.627 }_{- 0.622 }$ \\
$\log[E_\CUT$ (GeV)] & $ 0.371 ^{+ 0.100 }_{- 0.097 }$ & $ 0.328 ^{+ 0.147 }_{- 0.145 }$ & $ 0.209 ^{+ 0.091 }_{- 0.076 }$ & 
$ 0.314 ^{+ 0.123 }_{- 0.100 }$ & $ 35.299 ^{+ 106.386 }_{- 33.382 }$ & $ 0.043 ^{+ 0.170 }_{- 0.178 }$ \\
$\log[F_\gamma/F_\X$] & $ 3.450 ^{+ 0.307 }_{- 0.301 }$ & $ 2.489 ^{+ 0.598 }_{- 0.619 }$ & $ 0.471 ^{+ 0.300 }_{- 0.222 }$ & 
$ 1.100 ^{+ 0.577 }_{- 0.454 }$ & $ 27.647 ^{+ 98.743 }_{- 26.357 }$ & $ 0.962 ^{+ 0.684 }_{- 0.683 }$ \\
$\log[$CS] & $ 1.182 ^{+ 0.117 }_{- 0.110 }$ & $ 0.903 ^{+ 0.201 }_{- 0.199 }$ & $ 0.230 ^{+ 0.108 }_{- 0.101 }$ & 
$ 0.417 ^{+ 0.173 }_{- 0.144 }$ & $ 27.147 ^{+ 100.422 }_{- 25.653 }$ & $ 0.279 ^{+ 0.232 }_{- 0.221 }$ \\
$\log[$VI] & $ 1.684 ^{+ 0.048 }_{- 0.047 }$ & $ 1.700 ^{+ 0.055 }_{- 0.058 }$ & $ 0.087 ^{+ 0.049 }_{- 0.038 }$ & 
$ 0.105 ^{+ 0.059 }_{- 0.046 }$ & $ 3.377 ^{+ 4.634 }_{- 2.247 }$ & $ -0.016 ^{+ 0.074 }_{- 0.072 }$ \\
$\log[\Gamma]$ & $ 0.168 ^{+ 0.045 }_{- 0.052 }$ & $ 0.196 ^{+ 0.039 }_{- 0.052 }$ & $ 0.086 ^{+ 0.055 }_{- 0.042 }$ & 
$ 0.072 ^{+ 0.053 }_{- 0.040 }$ & $ 6.822 ^{+ 53.288 }_{- 5.822 }$ & $ -0.029 ^{+ 0.060 }_{- 0.060 }$ \\
$\log[$FWHM$ \cup \Delta_\gamma$] & $ -0.432 ^{+ 0.089 }_{- 0.101 }$ & $ -0.491 ^{+ 0.105 }_{- 0.121 }$ & $ 0.169 ^{+ 0.095 }_{- 0.082 }$ & 
$ 0.214 ^{+ 0.119 }_{- 0.097 }$ & $ 9.788 ^{+ 65.372 }_{- 8.777 }$ & $ 0.059 ^{+ 0.135 }_{- 0.131 }$\\
\enddata
\tablecomments{Statistical parameters obtained from the posterior 
distributions of the selected physical parameters: $p$ is the 
spin period, $\dot{p}$ the spin-down, $B_\LC$ the magnetic field 
strength at the light cylinder, $B_\SUR$ the surface magnetic field 
strength, $\dot{E}$ the spin-down power, $E_\CUT$ the spectral 
cutoff energy, $F_\gamma/F_\X$ the gamma-ray-to-X-ray flux ratio, 
CS the spectral curvature significance, VI the variability index, 
$\Gamma$ the photon index, and FWHM$ \cup \Delta_\gamma$ the union of 
full-width-half-maximum (FWHM) and peak separation ($\Delta_\gamma$). 
The error bars are the 99\% highest posterior density intervals 
(HPDIs). Since these values are obtained directly from the MCMC 
samplings, they represent the marginal distributions over the other 
statistical parameters. The values for $\mu^\RQ_X - \mu^\RL_X$ are 
directly measured from their marginal distributions.}
\end{deluxetable*}

\begin{figure*}
\gridline{\fig{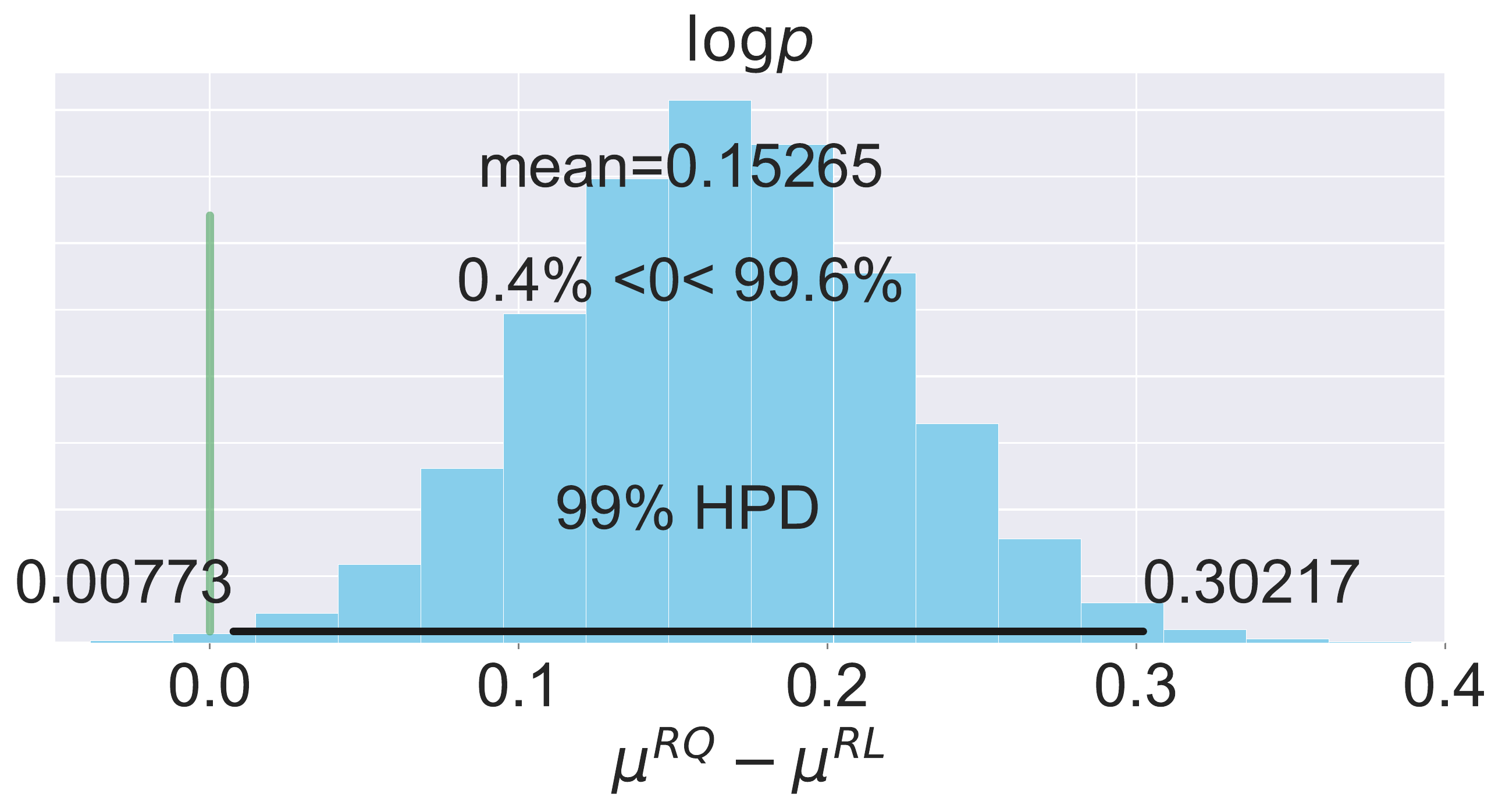}{0.45\textwidth}{}\fig{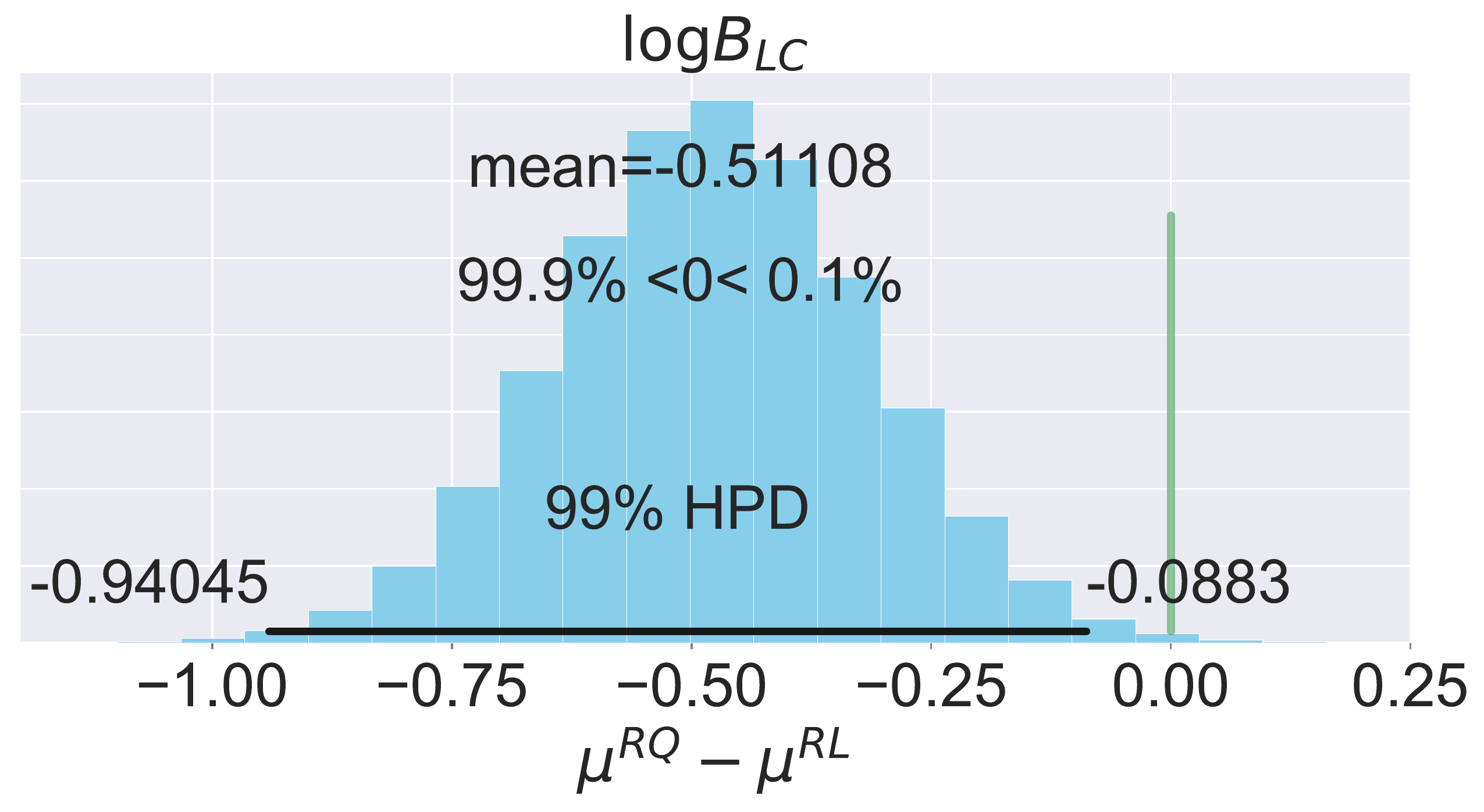}{0.45\textwidth}{}}
\gridline{\fig{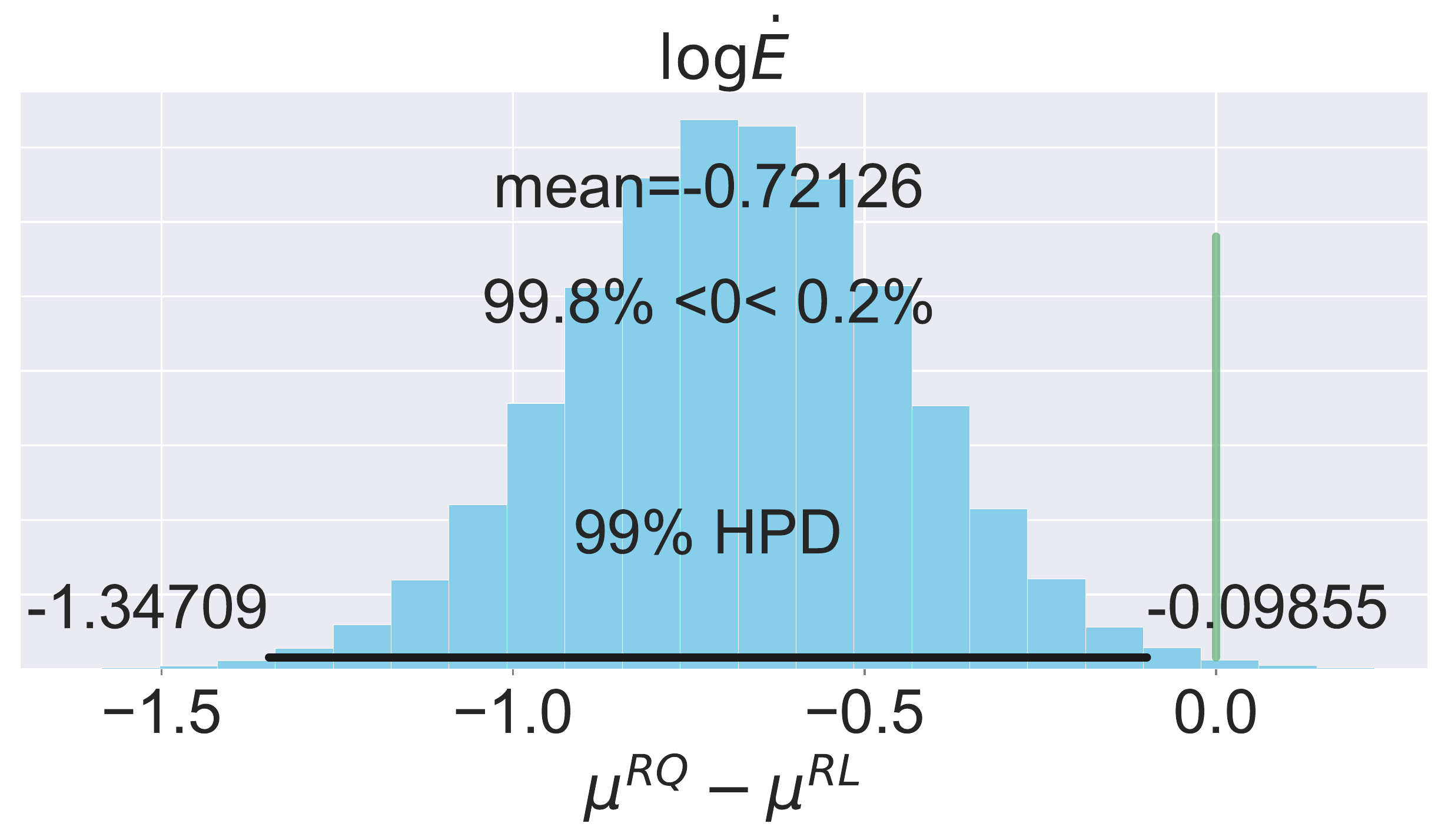}{0.45\textwidth}{}\fig{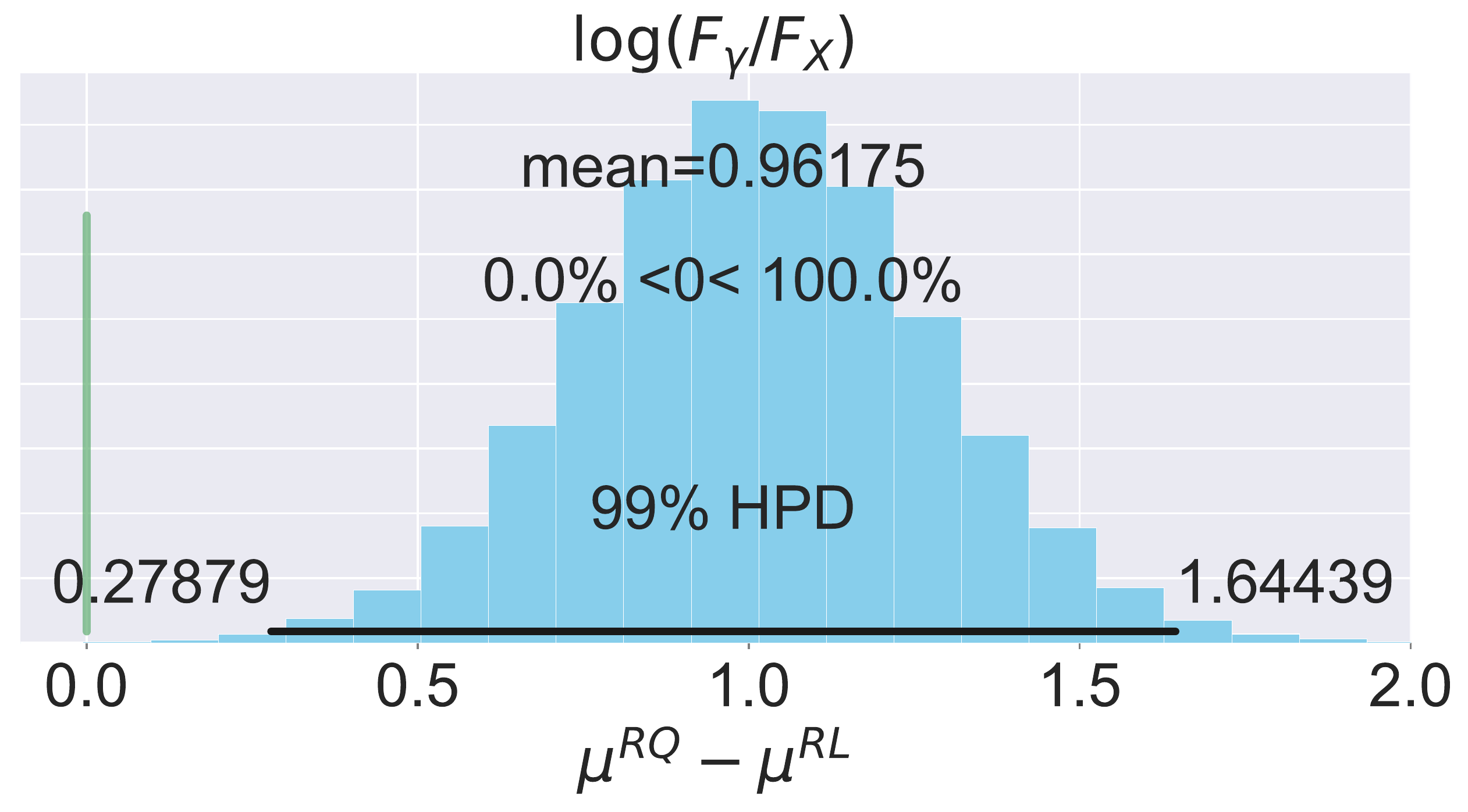}{0.45\textwidth}{}}
\gridline{\fig{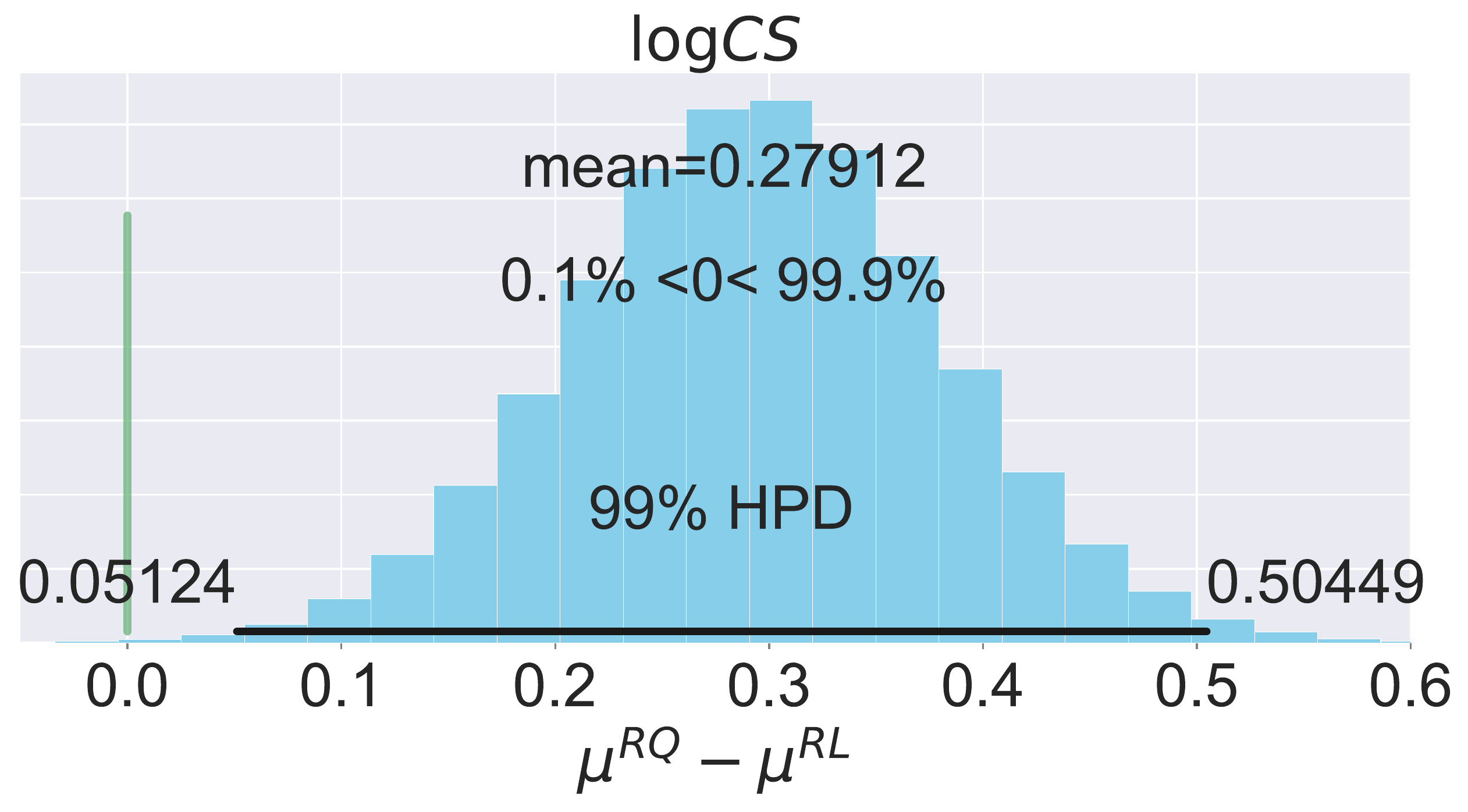}{0.45\textwidth}{}}
\caption{Marginal distributions of the difference of means 
between RQ and RL, $\mu^\RQ_X - \mu^\RL_X$, for the physical 
parameters $p$, $B_\LC$, $\dot{E}$, $F_\gamma/F_\X$, and 
and CS (from top left to bottom). The black horizontal lines
indicates the 99\% HPDIs with the lower and upper limits 
labeled. The percentages shown are the fractions of the 
distributions below and above $\mu^\RQ_X - \mu^\RL_X = 0$, 
indicated by the green vertical lines.\label{fig:diff_mu}}
\end{figure*}

\begin{figure*}
\gridline{\fig{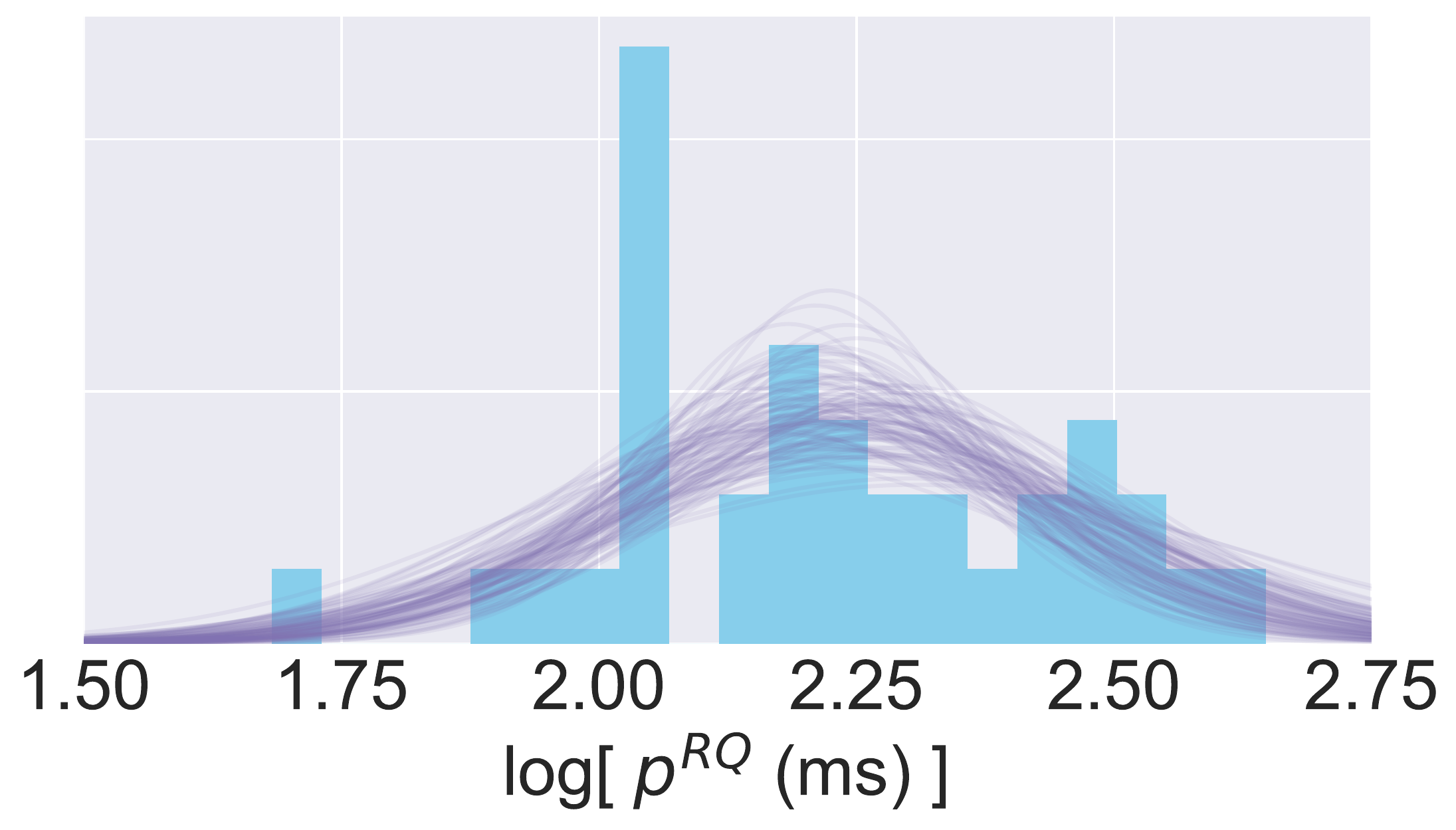}{0.45\textwidth}{}\fig{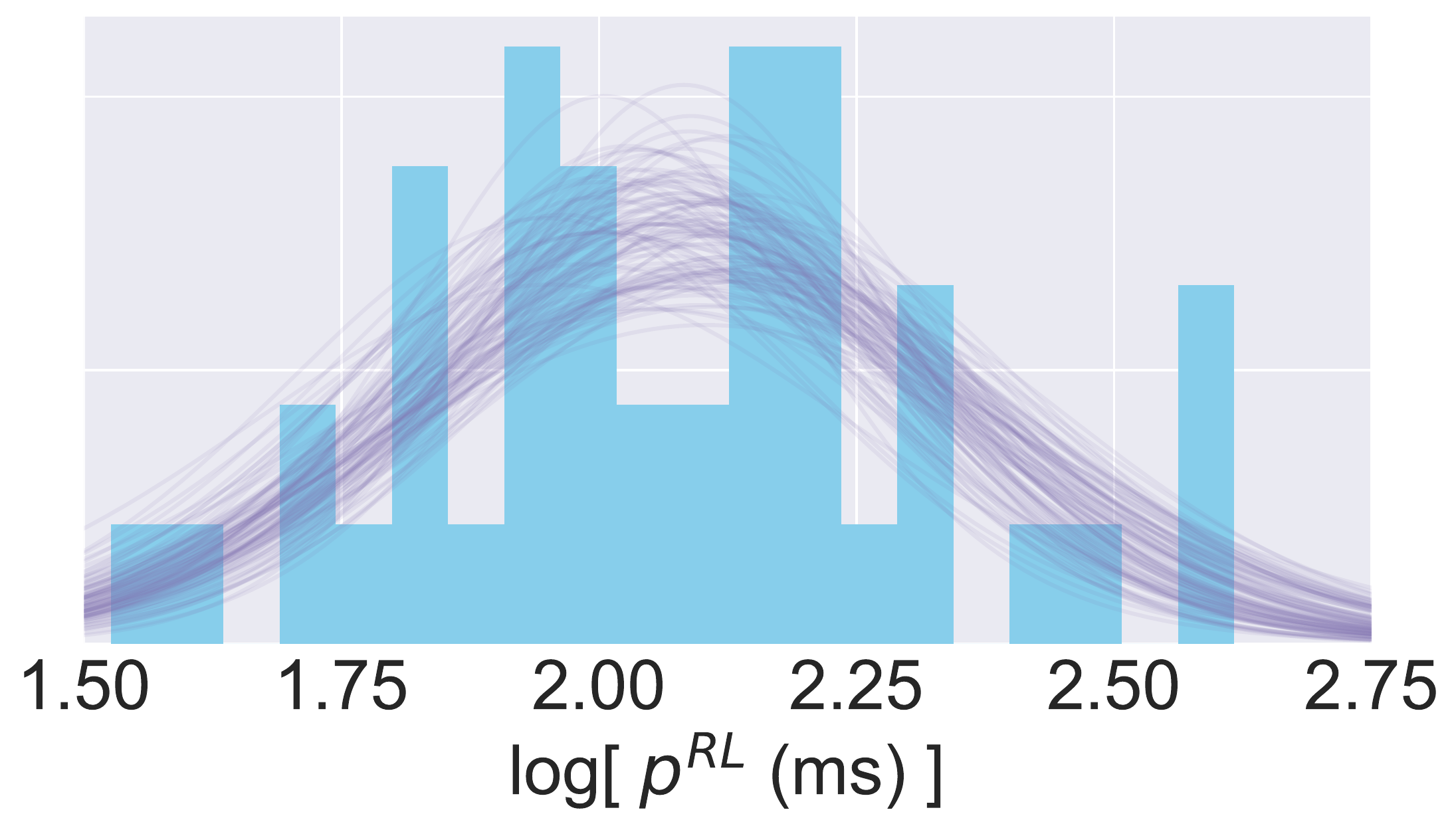}{0.45\textwidth}{}}
\gridline{\fig{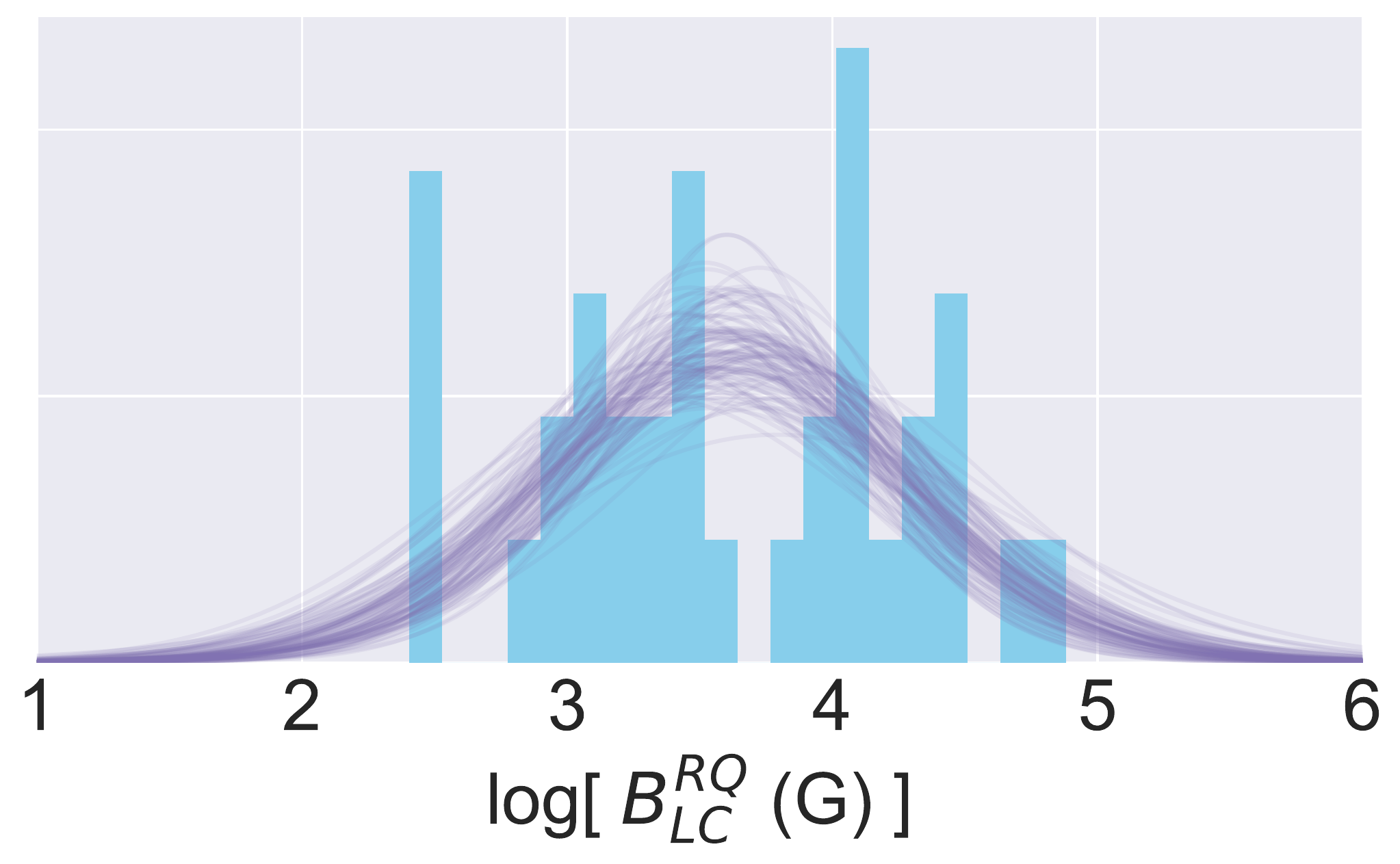}{0.45\textwidth}{}\fig{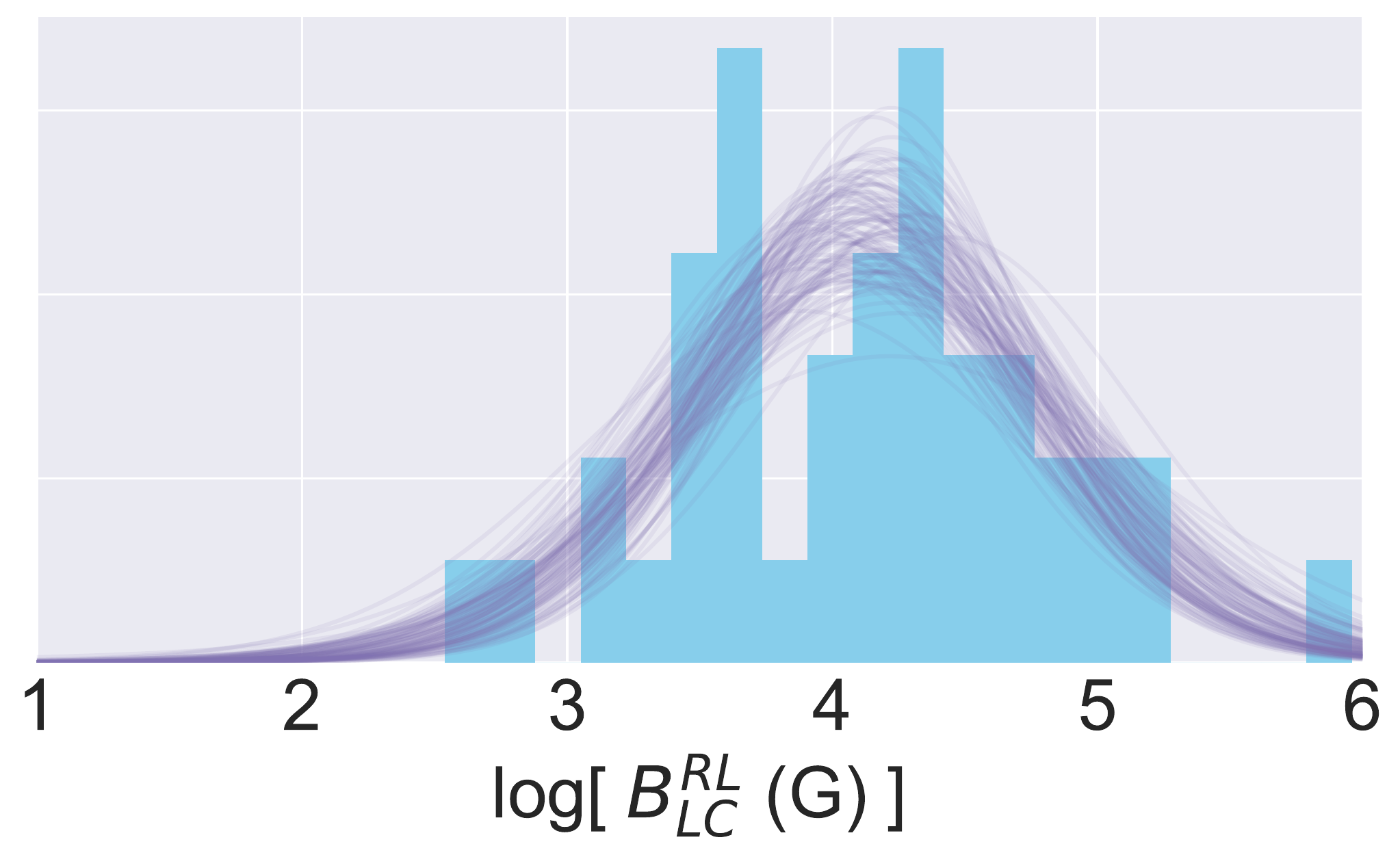}{0.45\textwidth}{}}
\gridline{\fig{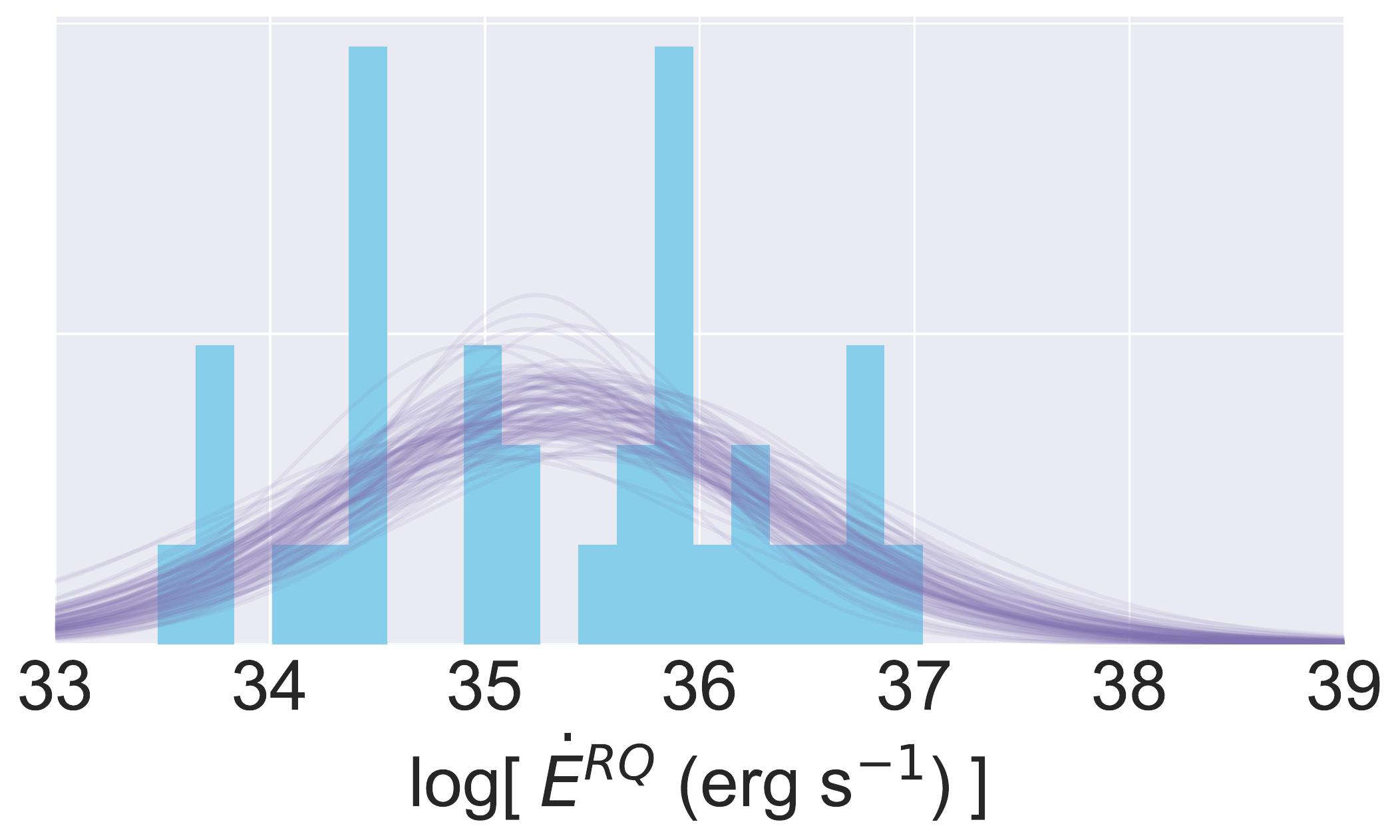}{0.45\textwidth}{}\fig{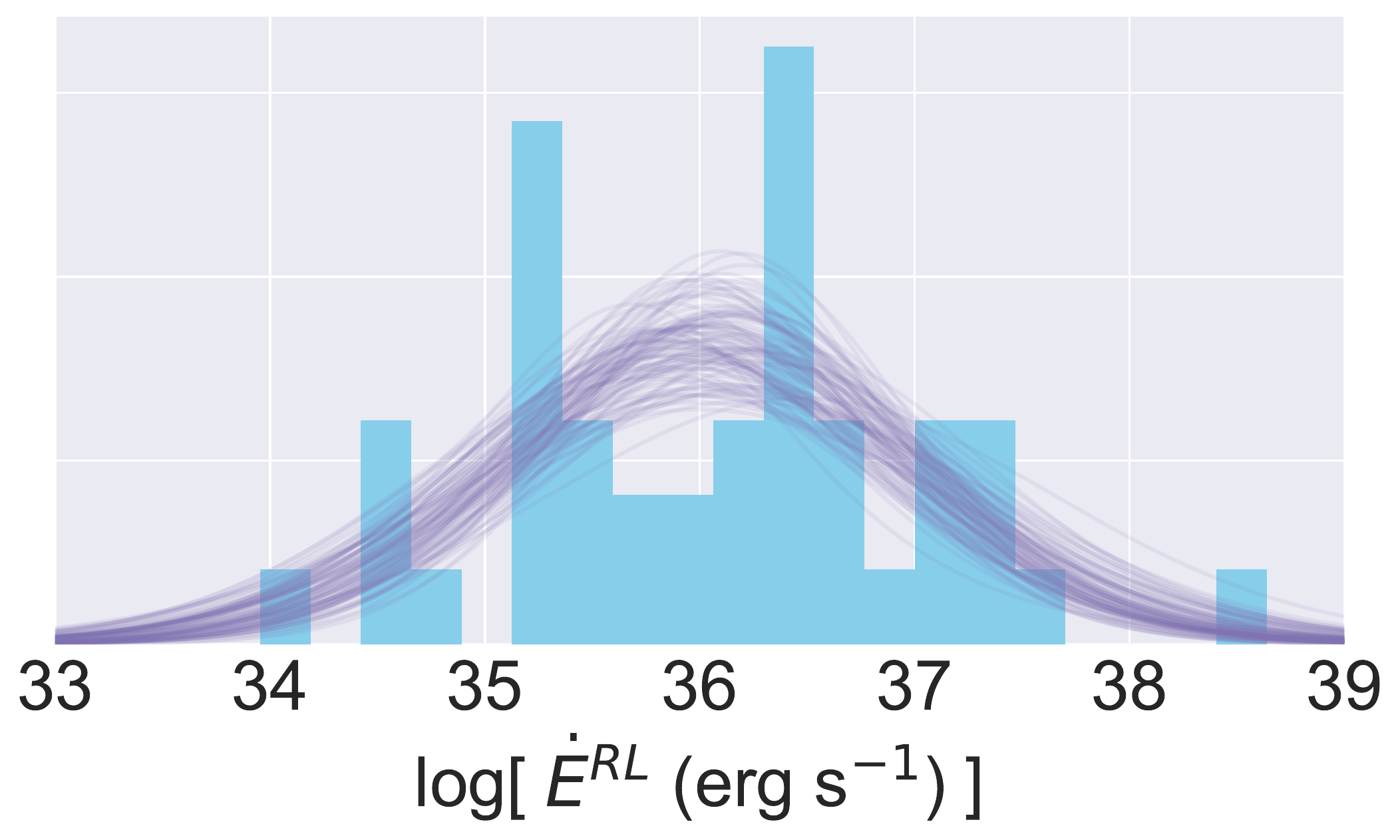}{0.45\textwidth}{}}
\caption{Observed distributions (histograms) with the 
posterior distributions (curves) from 100 randomly chosen 
traces overlaid for the RQ (left column) and RL (right column) 
distributions of $p$, $B_\LC$, and $\dot{E}$. For each 
physical parameter, the traces for RQ and RL are from the 
same 100 samplings.\label{fig:distri1}}
\end{figure*}

\begin{figure*}
\gridline{\fig{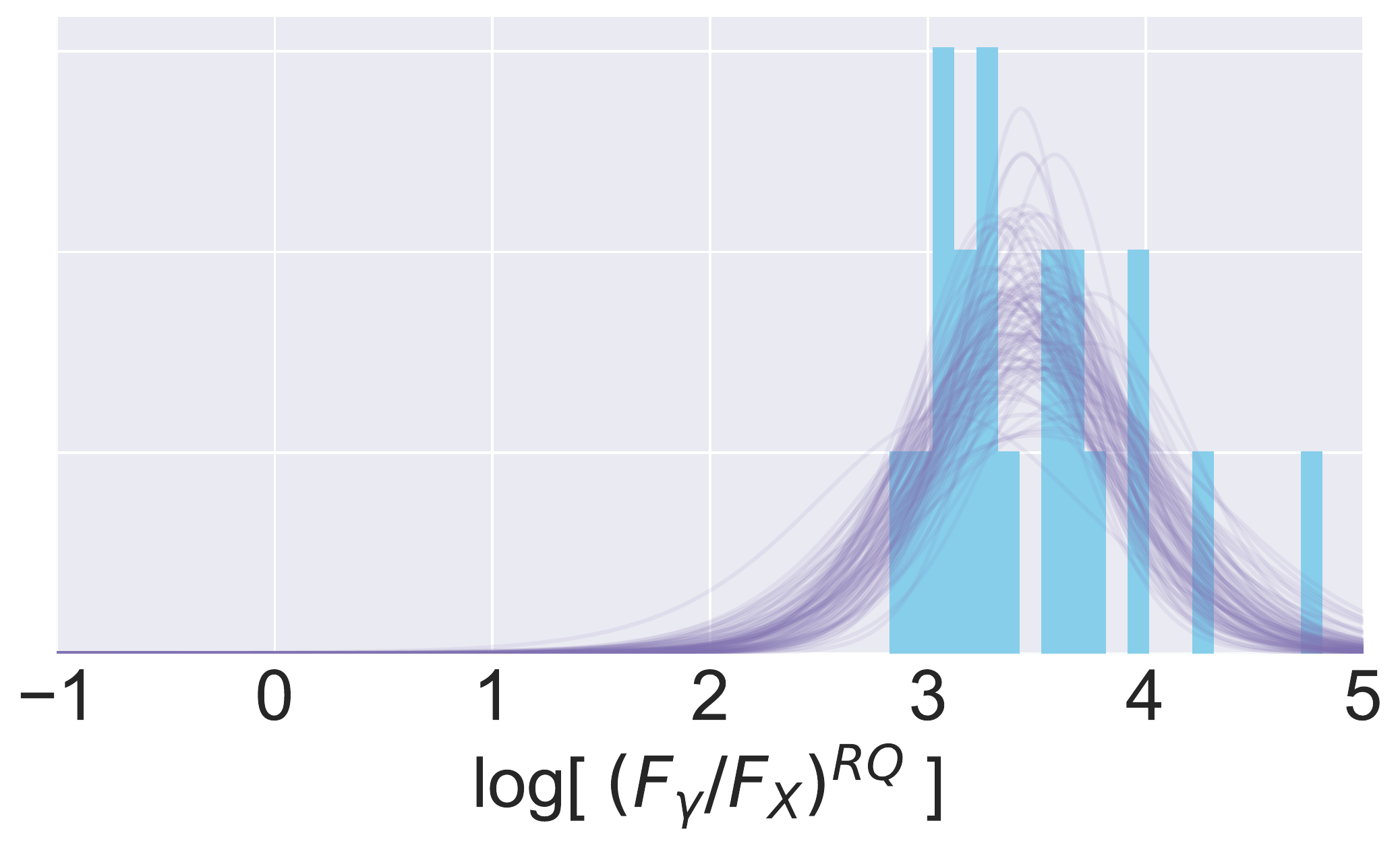}{0.45\textwidth}{}\fig{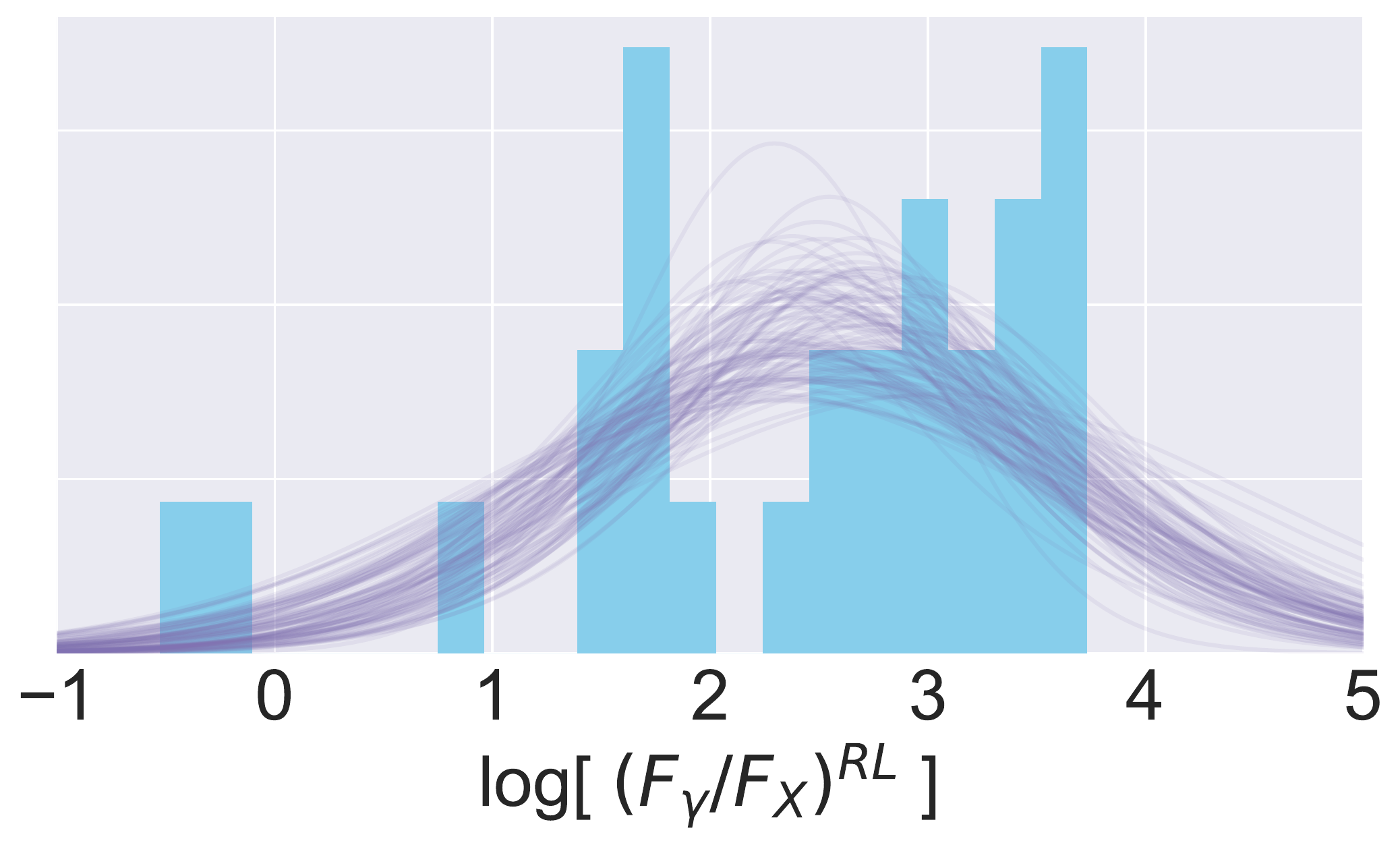}{0.45\textwidth}{}}
\gridline{\fig{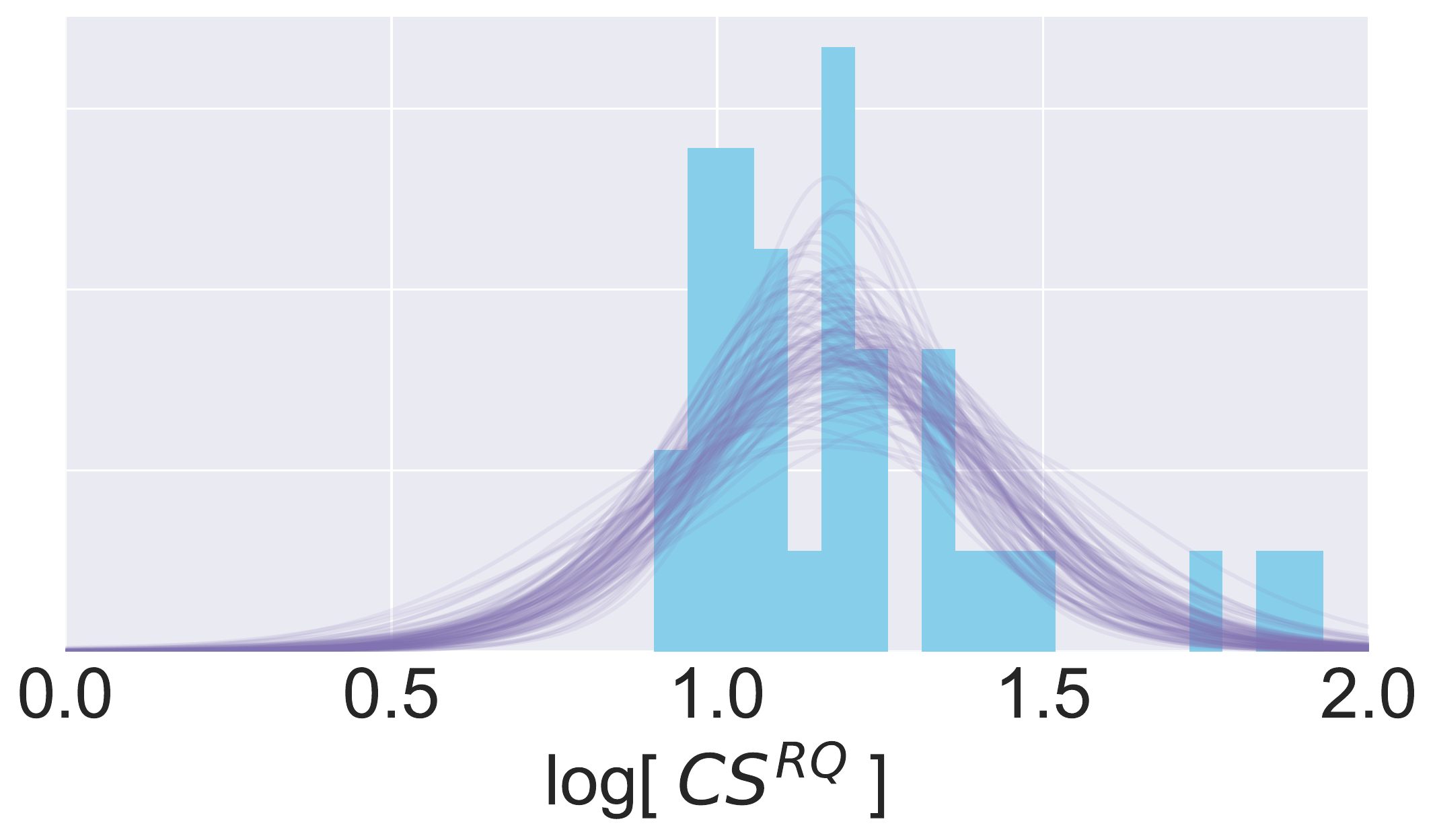}{0.45\textwidth}{}\fig{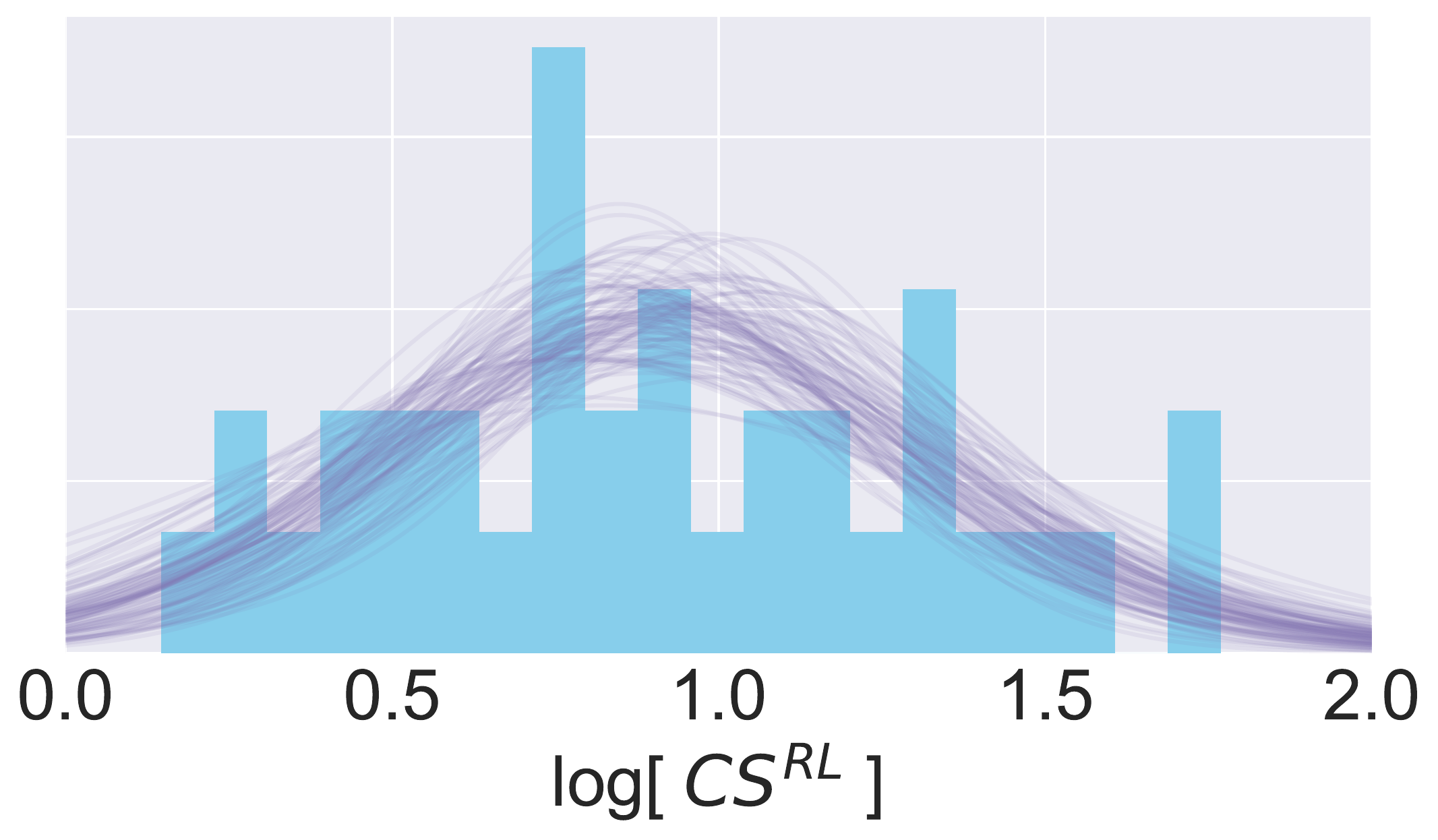}{0.45\textwidth}{}}
\caption{Same as Fig.~\ref{fig:distri1} for $F_\gamma/F_\X$ and CS.\label{fig:distri2}}
\end{figure*}

We apply the Bayesian Estimation Supersedes the t-Test (BEST) technique, 
developed by \citet{Kruschke13}, 
to the 12 observed distributions of the physical parameters listed 
in Tables 1 and 2 in H17. The analysis is performed using {\tt PyMC3} \citep{Salvatier16}. 
The BEST technique employs a non-standardised t-distribution as 
the likelihood function:
\begin{equation}
	\mathbb{P}(x|\mu ,\sigma ,\nu) = \frac{\Gamma\left(\frac{\nu+1}{2}\right)}{\Gamma\left(\frac{\nu}{2}\right)}\left(\frac{1}{\sqrt{\pi\mu}\sigma}\right)\left[1+\frac{1}{\nu}\left(\frac{x-\mu}{\sigma}\right)^2\right]^{-\frac{\nu+1}{2}},
\end{equation}
where $\mu$, $\sigma$, and $\nu$ are the statistical parameters of 
location, scale, and normality respectively, and $\Gamma$ is the gamma 
function. Since the t-distribution $\mathcal{T}(\mu ,\sigma ,\nu)$ 
is broader than a Gaussian distribution such that its variance is 
undefined, $\sigma^2$ is not the variance. Note that the statistical 
parameter $\nu$ is a measure of how heavy the probability is in the 
tails. If $\nu$ is large ($\nu \gtrsim 30$), the t-distribution 
approximates a Gaussian distribution.

The t-distribution is used because it can better account for the 
contributions from the  outliers in the population than a Gaussian 
distribution. Each physical parameter of RQ and RL pulsars is 
modeled with two t-distributions that are connected by the normality. 
Since the observed distributions are highly skewed in linear space, 
we take base-10 logarithm to all physical parameters, which means 
$\log X^\RQ \sim \mathcal{T}(\mu^\RQ_X ,\sigma^\RQ_X ,\nu_X)$ and 
$\log X^\RL \sim \mathcal{T}(\mu^\RL_X ,\sigma^\RL_X ,\nu_X)$. 
Therefore, we are essentially working with log-t-distributions.

The prior distributions of the statistical parameters are defined as
\begin{equation}
\begin{cases}
	\mu^\RQ_X &\sim \mathcal{N}(\bar{X},2S_X),\\
	\mu^\RL_X &\sim \mathcal{N}(\bar{X},2S_X),\\
	\sigma^\RQ_X &\sim \mathcal{U}(0.1S_X,10S_X),\\
	\sigma^\RL_X &\sim \mathcal{U}(0.1S_X,10S_X),\\
	\nu_X - 1 &\sim \mathcal{E}(1/29),
    \label{Eqn:priors}
\end{cases}
\end{equation}
where $\mathcal{N}$, $\mathcal{U}$, and $\mathcal{E}$ are the normal, 
uniform, and exponential probability distributions, respectively. We 
use $\bar{X}$ and $S_X$ to represent the mean and standard deviation 
of the physical parameter $X$, respectively, in order to distinguish 
from $\mu_X$ and $\sigma_X$, which have been defined earlier with 
different statistical meanings.

We perform a Markov chain Monte Carlo (MCMC) sampling using the 
above hierarchical Bayesian model. For each physical parameter 
we use 4 chains and obtain 25,000 traces per chain, resulting 
in posterior distributions consisting of 100,000 traces. The 
inferred statistical parameters for individual physical parameter 
are listed in Table~\ref{tab:posterior}.

We quantify the differences between the groups of RQ and RL 
pulsars by the distributions of $\mu^\RQ_X-\mu^\RL_X$. If 
$\mu^\RQ_X-\mu^\RL_X = 0$ is excluded from the 99\% highest 
posterior density interval (HPDI), then the two groups of 
pulsars are significantly different for this physical 
parameter. Among the 12 physical parameters listed in H17, 
we found that $\mu^\RQ_X-\mu^\RL_X$ for $X$ being the spin 
period $p$, the magnetic field strength at the light 
cylinder $B_\LC$, the spin-down power $\dot{E}$, the 
gamma-ray-to-X-ray flux ratio $F_\gamma/F_\X$, and the 
spectral curvature significance (CS) exhibit 99\% difference. 
The marginal distributions of $\mu^\RQ_X-\mu^\RL_X$ for 
these 5 parameters are shown in Fig.~\ref{fig:diff_mu}. 
Their observed and posterior distributions are plotted in Figs.~\ref{fig:distri1} and \ref{fig:distri2}.

We can check whether the Bayesian inference results make sense or not 
by the method of posterior predictive check (PPC). The PPC is a method 
to compare the predicted data generated by the posterior distributions 
to the observed data. We randomly choose 500 sets of the statistical 
parameters from the posteriors of the physical parameters, then we draw 
10,000 values from each of the 500 predicted data sets. We then 
calculate the mean of these 500 data sets and compare the posterior 
predictive distribution to the mean of the observed data. We found 
that the means are all centered at the predictive distributions, which 
implies they are consistent with each other. 

We also verified our results with different prior 
combinations for Eqn.~\ref{Eqn:priors} by uniform or normal 
distribution. In the most conservative case of using all 
uniform distributions, the posteriors of $\mu^\RQ_X$, 
$\mu^\RL_X$, $\sigma^\RQ_X$, and $\sigma^\RL_X$ are unchanged. 
In the case of $\nu_X$, it is unchanged if its value was small 
(i.e.,  outliers are important). Its posterior flavors large 
values for the cases that it was large ($\nu_X \gtrsim 30$), 
which means that the distribution is preferably described by 
a normal distribution and the shape is unchanged because a 
t-distribution approximates a normal distribution 
for $\nu_X \gtrsim 30$. 

\subsection{Fraction of radio-quiet gamma-ray pulsars} \label{subsec:frq}

\begin{figure}[h!]
\plotone{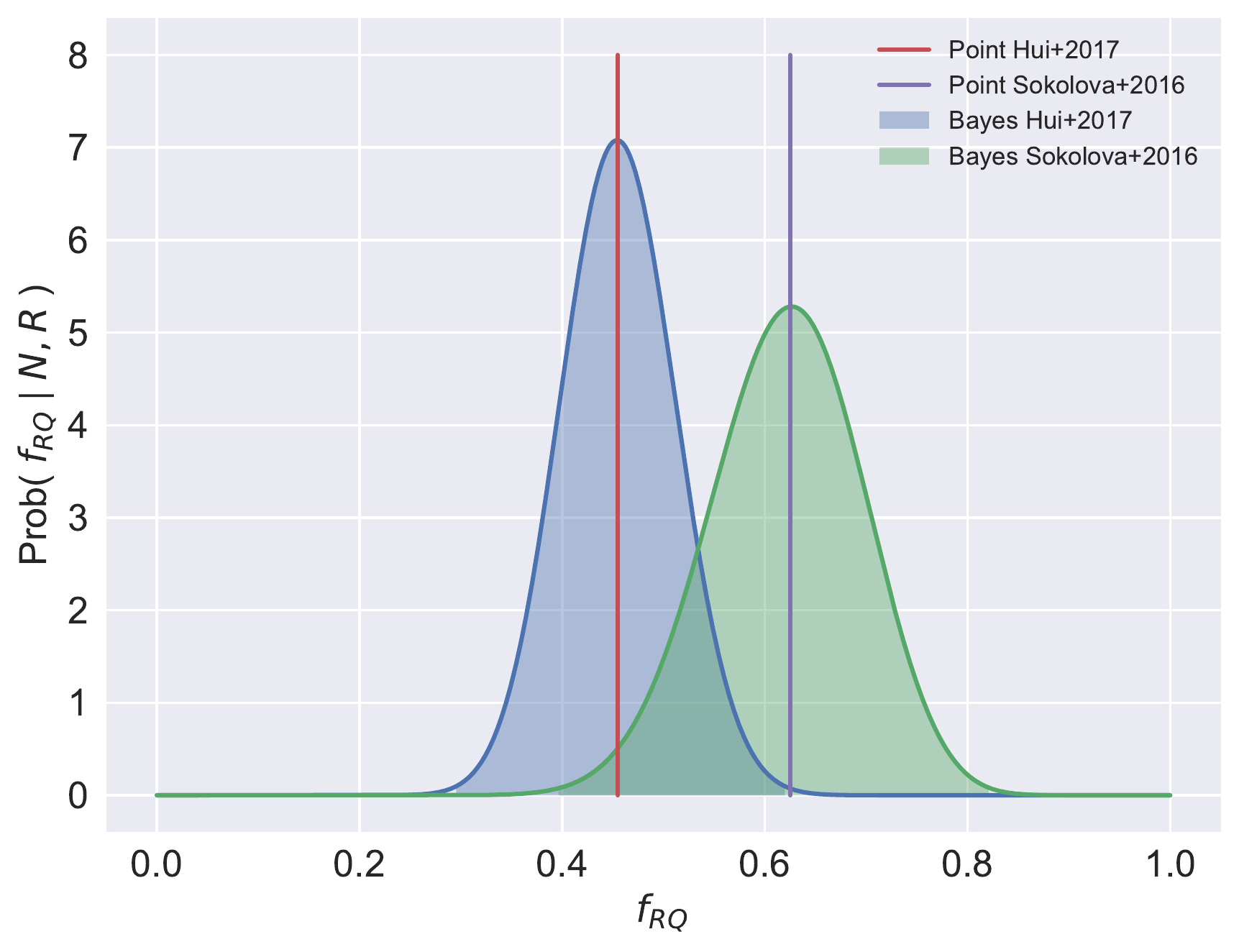}
\caption{Posterior distributions of the RQ fraction using the 
statistics from H17 (blue) and S16 (green). The shaded areas 
show the 99\% HPDIs. The vertical lines indicate the values 
of the point estimations for H17 (red) and S16 (purple). Note 
that the HPDI of S16 is wider than that of H17 due to fewer 
data points.
\label{fig:frq}}
\centering
\end{figure}

\begin{figure*}
\gridline{\fig{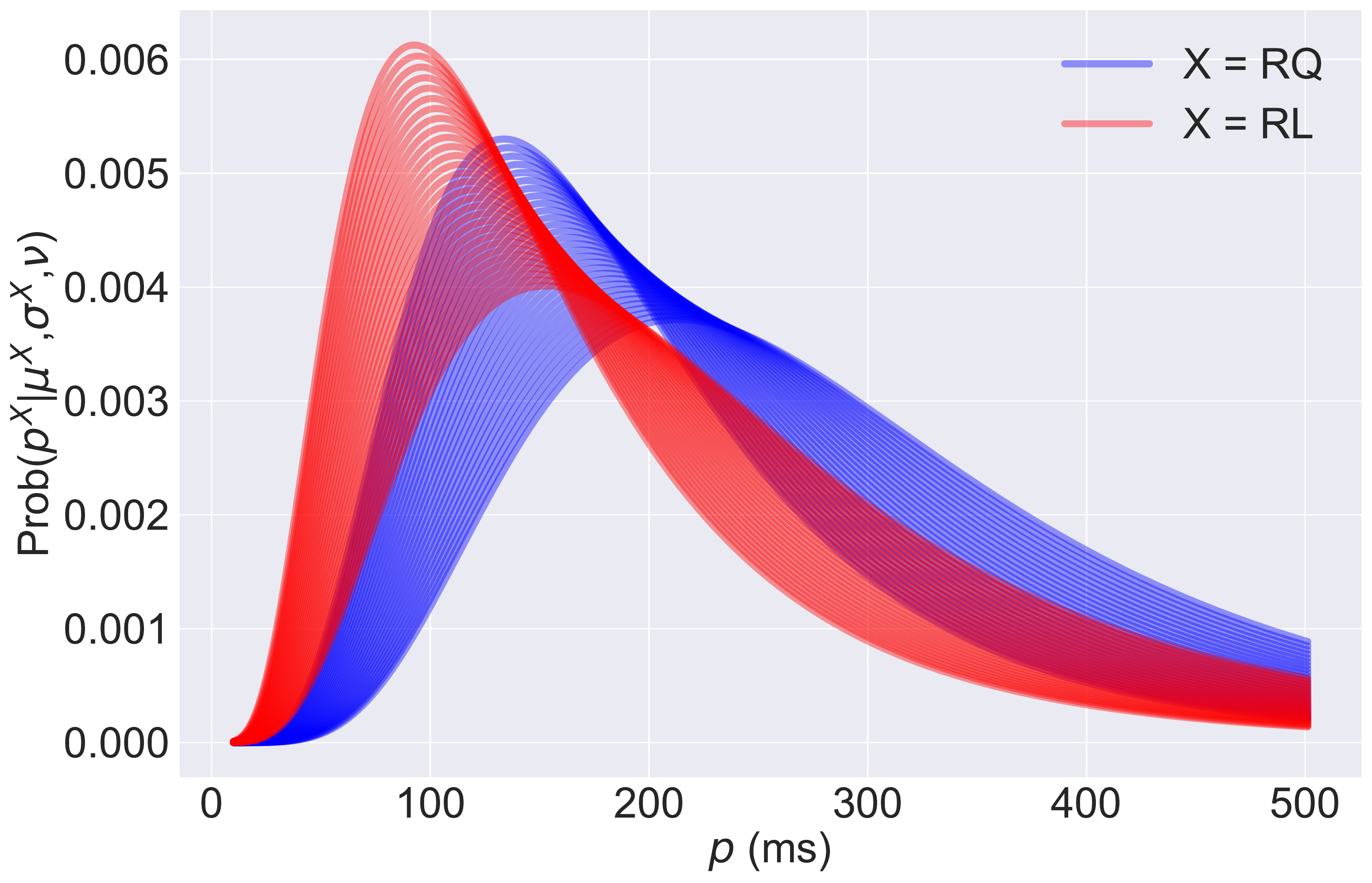}{0.5\textwidth}{(a)}
          \fig{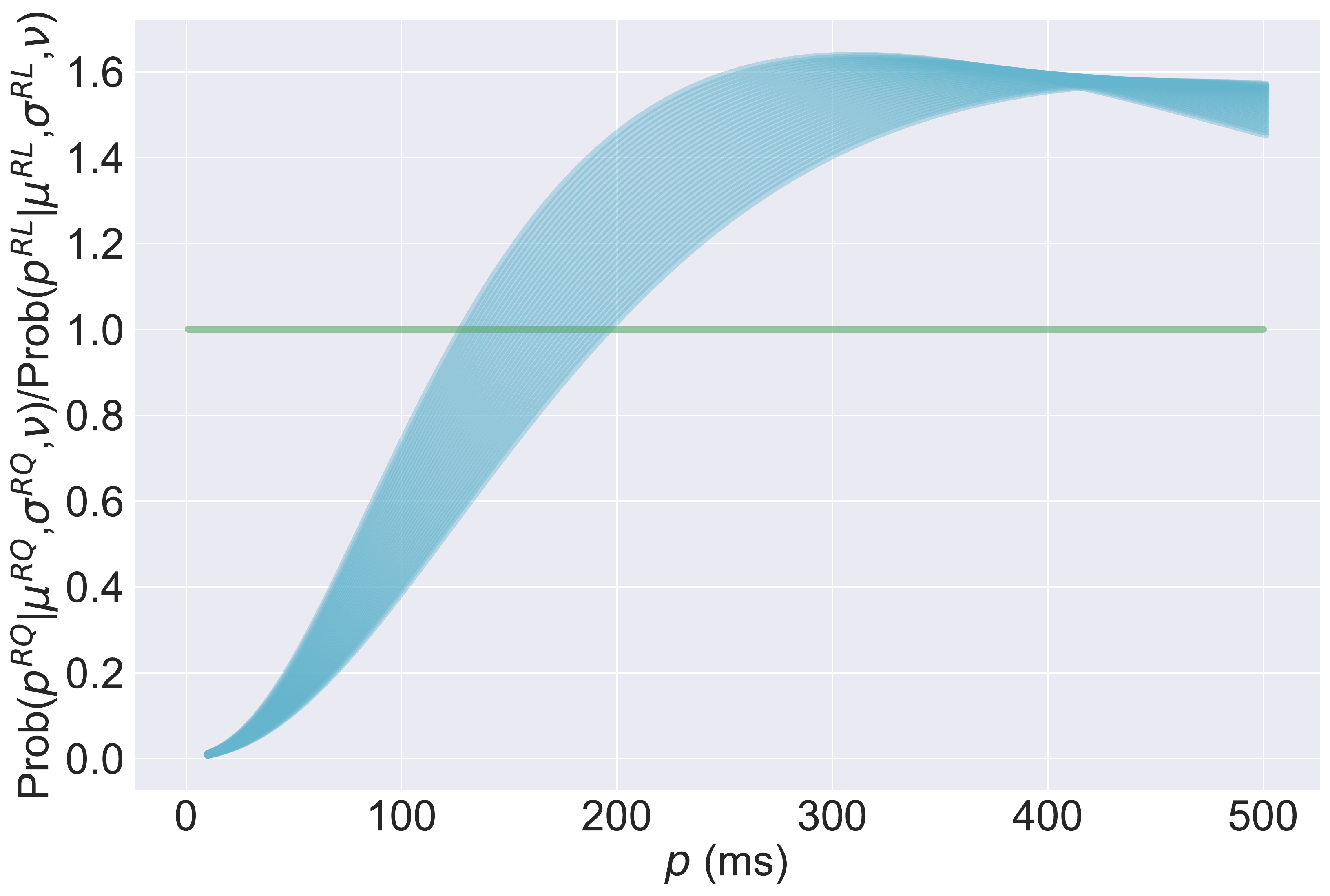}{0.5\textwidth}{(b)}
          }
\caption{(a) Posterior distributions of $p^\RQ$ and $p^\RL$ obtained
from the mean values of the statistical parameters of the posteriors. 
The shaded regions are plotted using the 99\% HPDIs of $\mu^\RQ$ 
and $\mu^\RL$. (b) Relative numbers of RQ to RL pulsars expected to 
be detected given a value of $p$. The green line is $y=1$.\label{fig:Prob_p}}
\end{figure*}

The detections of $R$ RQ pulsars and $(N-R)$ RL pulsars, 
where $N$ is the total number of pulsars, form a binomial 
distribution with success probability as the fraction of 
RQ pulsars $f_\RQ$. Thus the likelihood of detecting $R$ 
RQ pulsars among $N$ pulsars is 
\begin{equation}
	\mathbb{P}(N,R|f_\RQ) \propto (f_\RQ)^R \times (1 - f_\RQ)^{N-R}.
\end{equation}
The equal sign is replaced by the proportional sign because the binomial 
coefficient can be absorbed into the normalisation of the posterior distribution.

We use the non-informative Jeffreys prior for the 
binomial distribution
\begin{equation}
	\mathbb{P}(f_\RQ) = \frac{1}{\sqrt{f_\RQ(1 - f_\RQ)}},
\end{equation}
which is indeed a beta distribution 
Beta$(\frac{1}{2},\frac{1}{2})$. The Jeffreys prior 
represents the fact that we are ignorant about the 
probability that the model is true (in our case, 
that $f_\RQ$ equals certain value), and it has the 
property that it is independent of the model parameterisation.

The posterior distribution $\mathbb{P}(f_\RQ|N,R)$ using 
the data from H17 is displayed in Fig.~\ref{fig:frq}. 
For comparison, the posterior using the data from S16 is 
also displayed. The frequentist point estimations, 
$f_\RQ = 35/77 = 0.455$ for H17 and $f_\RQ = 25/40 = 0.625$ 
for S16, are consistent with the 99\% HPDIs of the 
posteriors, $f_\RQ = 0.454^{+0.166}_{-0.159}$ for H17 and 
$f_\RQ = 0.627^{+0.195}_{-0.230}$ for S16. Note that our result 
indicates that the inferred fraction of RQ pulsars from H17 and S16 
are indeed consistent to each other, because their HPDIs overlap.

We can further calculate the expected ratio of RQ to RL pulsars 
to be detected given a value of $p$. The posterior distributions 
of $p^\RQ$ and $p^\RL$ obtained from Sect.~\ref{subsec:best} by 
taking the mean values of the distributions of $\mu^\RQ_p$ and 
$\mu^\RL_p$ are displayed in Fig.~\ref{fig:Prob_p}(a). It is 
observed that $RQ$ pulsars are usually found with larger $p$. 
This can be seen more clearly if we plot the relative numbers 
of RQ to RL pulsars expected to be detected given a value of 
$p$ in Fig.~\ref{fig:Prob_p}(b). The plot indicates that the 
probabilities of detecting a RQ and RL pulsar is 50/50 at 
$p \approx 120\textsc{-}200$~ms.

\subsection{Constraining the radio-cone opening half-angle and magnetic inclination angle} \label{subsec:angle}

If we assume (1) the gamma-rays are emitted from the outer 
gap \citep[e.g.,][]{Cheng98,Takata06,Takata08}, (2) the radio 
emissions originate from the polar cap region, and (3) there 
is no intrinsic physical difference between RQ and RL pulsars, 
then the detection of radio signals depends completely on 
geometrical effects. Since all gamma-ray pulsars are detected 
in gamma-rays by definition, we know that the LOS 
from the Earth must cut through the outer gap. Therefore we can 
replace the random variable $f_\RQ$ in Sect.~\ref{subsec:frq}
with the radio-cone opening half-angle, $\delta$, and the 
magnetic inclination angle, $\alpha$. The fraction $1 - f_\RQ$ 
is just the solid angle swapped by the radio cone over 2$\pi$. 
Using simple geometry, we have
\begin{equation}
	f_\RQ(\alpha,\delta) = \begin{cases}
	\cos(\alpha+\delta) &\text{for $\alpha < \delta$},\\
    1-2\sin\alpha\sin\delta &\text{for $\delta \leqslant \alpha \leqslant \frac{\pi}{2}-\delta$},\\
    1-\cos(\alpha-\delta) &\text{for $\alpha \geqslant \frac{\pi}{2}-\delta$},
	\end{cases}
\end{equation}
and the likelihood of detecting $R$ RQ pulsars 
among $N$ pulsars is
\begin{equation}
	\mathbb{P}(N,R|\alpha,\delta) \propto [f_\RQ(\alpha,\delta)]^R \times [1-f_\RQ(\alpha,\delta)]^{N-R}.
\end{equation}

It is expected that the spinning axis and the magnetic axis are 
randomly placed at the moment when the pulsar was born \citep[e.g.,][]{Rookyard15b}. 
From symmetry, it is sufficient to only consider a single hemisphere, 
and the opening angle of the radio cone cannot be larger than 
$180^{\circ}$, so that we have $(\alpha,\delta) \in [0,\frac{\pi}{2}]$. 
Therefore we use a uniform prior for $\alpha$:
\begin{equation}
	\alpha \sim \mathcal{U}(0,\frac{\pi}{2}).
\end{equation}
An empirical relation $\delta \propto p^{-n}$ is found by 
past observations \citep[e.g.,][]{Narayan83,Lyne88,Biggs90,Gil93,Gil96},
where $n \approx 1/2$. We adopt the relation 
\begin{equation}\label{eqn:angles_rel}
	\delta = 6.3^{\circ} \times p^{-1/2}
\end{equation}
from \citep{Gil96} at 1.4~GHz for the outer beam. In order 
to obtain a reasonable $\delta$ prior, we repeat the BEST 
technique on the whole population $p = p^\RQ \cup p^\RL$ 
using the same procedure described in Sect.~\ref{subsec:best}, 
and obtained the posterior distribution for $\log p$. 
Using Eqn.~(\ref{eqn:angles_rel}), we can construct the 
prior of $\delta$ using this t-distribution 
\begin{equation}
	\log p \sim \mathcal{T}(\mu_p ,\sigma_p ,\nu_p).
\end{equation}

\begin{figure*}
\gridline{\fig{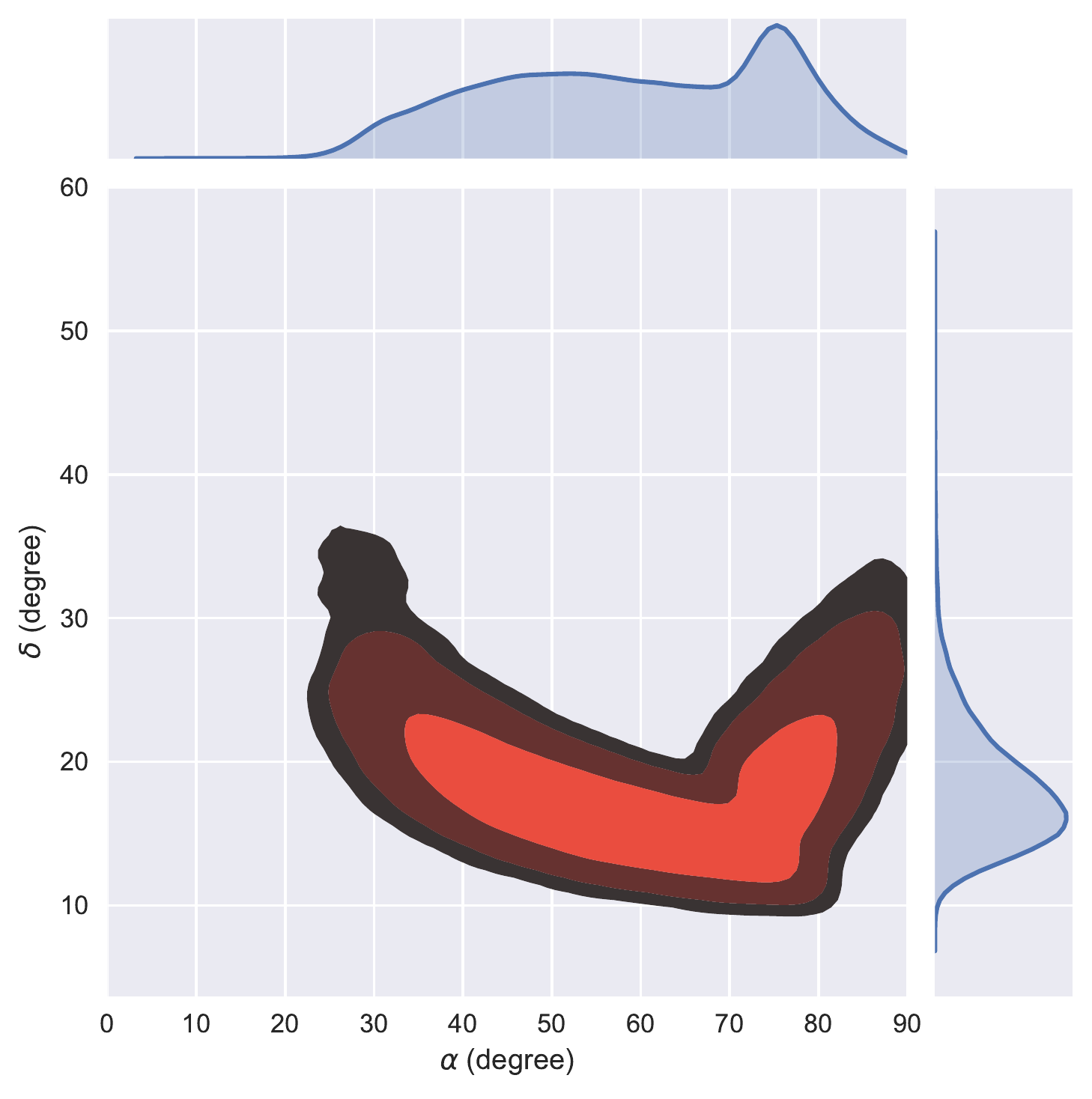}{0.5\textwidth}{(a)}
          \fig{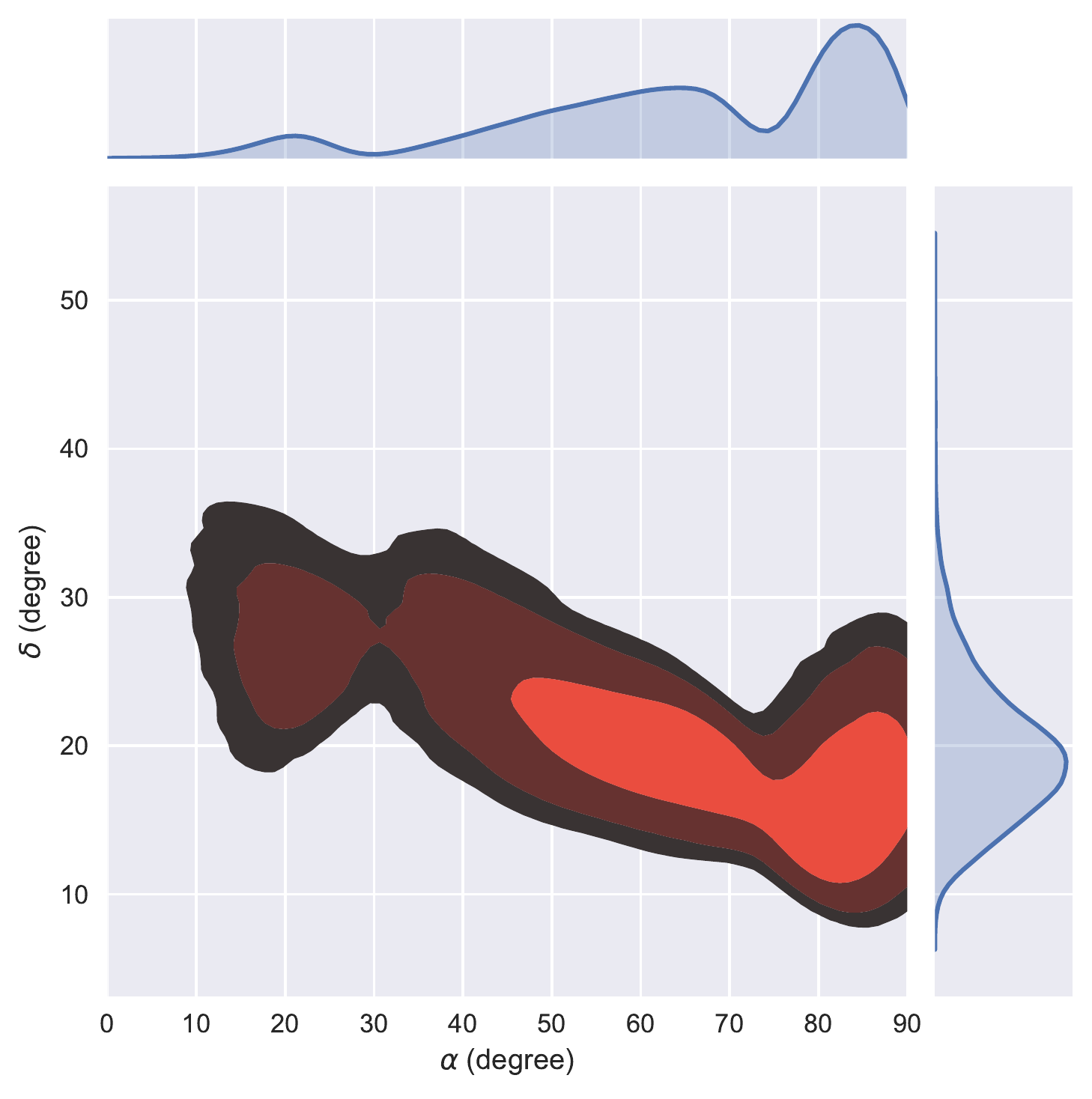}{0.5\textwidth}{(b)}
          }
\caption{Two-dimensional posterior distributions of 
$\alpha$ and $\delta$ obtained from (a) H17 and (b) S16. 
The shaded regions are the 67\%, 95\%, and 99\% HDPIs. 
The marginal distributions for $\alpha$ and $\delta$ 
are also plotted at the top and right of each panel,
respectively.\label{fig:angles_post}}
\end{figure*}

The posterior of this hierarchical Bayesian model is
a function of $\alpha$ and $p$ (or $\delta$), 
conditioning on $R$ and $N$. From the values of $R$ 
and $N$ from H17 and S16, we can compute the posteriors 
$\mathbb{P}(\alpha,\delta|N,R)$ for both studies, which 
are plotted in Fig.~\ref{fig:angles_post}. The results 
suggest, based on the current detection statistics of 
RQ and RL pulsars, assuming Eqn.~(\ref{eqn:angles_rel}), 
that $\alpha$ is skewed towards values larger than $40^{\circ}$, 
and $\delta$ is between $10^{\circ}$ and $35^{\circ}$.

It may be interesting to remove the dependence of $p$ 
in $\delta$ by inferring $\delta \times p^{1/2}$. 
However, $p$ is not uniquely determined but a random 
variable, it is not possible to obtain the distribution 
for $\delta \times p^{1/2}$ since $\delta$ depends only 
on $p$. The distribution of $\delta$ is given once that 
of $p$ is given. To verify this, we repeated the analysis 
using $\delta \times p^{1/2}$ instead of $\delta$, which 
results in a point estimation corresponding to $6.3^{\circ}$.

\section{Conclusions and Discussions} \label{sec:disc}

We demonstrated for the first time using Bayesian statistics 
to study the populations of RQ and RL gamma-ray pulsars. 
The analysis results are robust to  outliers in the pulsar 
populations. The Bayesian approach has advantages over the 
conventional frequentist approach when dealing with scarce 
number of data points, as in the current case of RQ and RL 
gamma-ray pulsars. We obtained complete information of the 
posterior distributions for every physical and statistical 
parameters under study, in the sense that the uncertainties, 
modality, and correlations between parameters are readily 
visualised.

We showed that the spin period $p$, the magnetic field strength 
at the light cylinder $B_\LC$, the spin-down power $\dot{E}$, 
the gamma-ray-to-X-ray flux ratio $F_\gamma/F_\X$, and the spectral 
curvature significance CS of the two groups of pulsars exhibit 
significant differences at the 99\% level. From 
Fig.~\ref{fig:diff_mu} it can be seen that the RQ pulsars have 
$p$, $F_\gamma/F_\X$, and CS larger than those of the RL pulsars, 
while $B_\LC$ and $\dot{E}$ of RQ pulsars are smaller than RL 
pulsars. These behaviours can be explained because both $B_\LC$ 
and $\dot{E}$ scale with $p^{-3}$ \citep{Hui17}.

To investigate further the difference in $F_\gamma/F_\X$, we 
repeat the BEST analysis on the gamma-ray and X-ray luminosities, 
$L_\gamma$ and $L_\X$, and the radiation efficiencies of gamma 
rays and X rays, $L_\gamma/\dot{E}$ and $L_\X/\dot{E}$, 
respectively, in base-10 logarithmic scale. For $L_\gamma$ and $L_\X$, 
we found that the $L_\X$ populations of RQ and RL pulsars are 
different at the 99\% level, while that of $L_\gamma$ has 
no significant difference. However, for $L_\gamma/\dot{E}$ and 
$L_\X/\dot{E}$, we found that $L_\gamma/\dot{E}$ of RQ and RL 
pulsars are different at the 95\% level, while that of 
$L_\X/\dot{E}$ has no significant difference. Therefore, the 
observed difference in $F_\gamma/F_\X$ may be a mixed effect 
on the luminosities and spin-down powers in the gamma-ray and 
X-ray bands. Note that this effect is also subjected to the 
uncertainties of the adopted distances. 

We showed that the averaged spin period $p$ of the RL pulsars 
is smaller than that of the RQ pulsars. This result can be 
understood if the average width of the radio cone of RL pulsars 
is larger than that of the RQ pulsars. As we adopted in equation~(\ref{eqn:angles_rel}),
it has been considered that the width of the radio beam is 
related to the size of the polar cap, whose radius is typically 
$R_\POLAR \sim R_\NS(R_\NS/R_\LC)^{1/2}\propto p^{-1/2}$, where 
$R_\NS$ is the radius of the neutron star and $R_\LC=pc/2\pi$ 
is the radius of the light cylinder. It has also been suggested 
that the radio emission altitude relative to the radius of the 
light cylinder increases with decreasing $p$, i.e., the emission 
altitude approaches to the light cylinder for smaller $p$ \citep{Kijak03}. 
These empirical relations suggest that the width of the radio cone 
increases with decreasing $p$, and hence the pulsar with a shorter 
period has a large chance to be detected as a RL pulsar. This 
explains that the average $p$ of the RL pulsars is larger than 
that of the RQ pulsars. Since $B_\LC \propto B_\SUR p^{-3}$ 
and $\dot{E} \propto B_\SUR^2 p^{-4}$, the average $B_\LC$ and 
$\dot{E}$ of the RL pulsars are larger than that of the RQ pulsars.

The gamma-ray spectral curvature measured by the Fermi-LAT rules 
out the classical polar cap scenario \citep[e.g.,][]{Daugherty96} 
and supports the hypothesis that the gamma-ray emission is 
originated from the outer magnetosphere, e.g., \citet{Arons83} 
for the slot gap model, \citet{Cheng86} for the outer gap model, 
and \citet{Spitkovsky06} for the current sheet. However, the 
spectral behavior above the cutoff energy at around 3~GeV, 
which decays slower than pure exponential function \citep{Ackermann12}, 
has not been fully understood. It has been discussed that the 
spectral curvature is related to the non-stationary activity of 
the outer gap emission \citep{Takata16} or the magnetic field 
structure around the light cylinder \citep{Vigano15}.

To account for the difference of the gamma-ray spectral curvature 
between the RL and RQ pulsars, we may expect another component 
that contributes to the high-energy emissions for the RL pulsars. 
For example, we speculate that inverse Compton (IC) process plays 
a role in the high-energy photon production of the RL pulsars. 
For the RL pulsars, which generally have wider radio cones and 
have emission regions closer to the light cylinder than their 
RL counterparts, it is highly plausible that part of the radio 
waves get into the acceleration region around the light cylinder. 
Since typical Lorentz factor of the primary electrons/positrons 
is $\Gamma \sim 3 \times 10^7$, IC scattering of the primary 
electrons/positrons with radio waves ($\sim 1$~GHz) will 
produce photons in the GeV regime. On the other hand, 
the probability of radio photons from the RQ pulsars getting 
into the gap is low. Hence the spectral curvature of the RQ 
pulsars could be larger than that of the RL pulsars. 

While the observed difference of the spectral curvature can be 
explained by the IC scenario, this can also be stemmed from 
observational bias. Different from the RL cases, RQ pulsars can 
only be detected through blind search. This leads to the fact 
that the signal-to-noise ratio of RQ pulsars are higher on 
average. For RL pulsars, with the help of the radio ephemeris, 
fainter pulsars can also be detected. For these fainter sources, 
the cutoff may not be constrained. This might lead to an 
apparently less curved spectra.

The differences between the results using the data from H17 and 
S16 arise from the fact that they used different pulsar samples. 
However, we found that their Bayesian inferred fractions of RQ 
pulsars are indeed consistent with each other. H17 adopted all 
non-recycled gamma-ray pulsars from 2PC and 3FGL, while S16 
employed a blind search for gamma-ray pulsars using the 
{\it Fermi} data alone, resulting in fewer number of data points 
but plausibly bias-free. 

From Fig.~\ref{fig:angles_post}(b) it is observed that the 
$\alpha$-$\delta$ posterior distribution of S16 is multi-modal, 
and extends to lower vales of $\alpha$. This is because S16 has 
fewer data points but a higher fraction of RQ pulsars. Since the 
ranges of plausible values of $\delta$ are similar for H17 and S16, 
smaller $\alpha$ results in smaller solid angle on the sky, i.e., 
more RQ pulsars. Moreover, it can be seen that the posterior of S16 
is more heavily distributed in regions where $\alpha + \delta > 90^\circ$. 
The coverage of the radio cone increases with $\alpha$ and $\delta$, 
but decreases when $\alpha + \delta > 90^\circ$ because the 
projections of the bipolar cones on opposite hemispheres overlap. 
This again implies higher fraction of RQ pulsars than RL pulsars.

Using Bayesian inference, assuming pure geometrical effects, 
we showed that the distribution of the magnetic inclination 
angle $\alpha$ is skewed towards larger values, which does 
not exhibit the reported unexpected skew of $\alpha$ 
towards smaller values from \citet{Rookyard15a}, and is 
consistent with the conclusion of \citet{Rookyard15b} 
that $\alpha$ should be skewed towards larger values. On the 
other hand, the values of the radio-cone half-angle $\delta$ 
are constrained to be within $10^\circ$ and $35^\circ$. 
Since we used equation~(\ref{eqn:angles_rel}) and considered 
pulsars with rotation period within 0.03~s and 0.2~s, we 
should obtain $\delta$ to be within $10^{\circ}$ and $35^{\circ}$. 
However, using Bayesian statistics instead of conventional 
frequentist statistics gives us not only a range estimation 
for $\delta$, but also for $\alpha$ and their joint posterior, 
which gives their uncertainties and correlations for free. 
From the gamma-ray emission point of view, larger inclination 
angles are preferentially detected by the observations 
\citep[c.f.,][]{Watters11,Takata11}.

In the future, the Bayesian inference results shown here could be
updated when more RQ and RL gamma-ray pulsars are detected. The 
uncertainties (expressed in HPDIs) will be reduced because the 
observed distributions in Figs.~\ref{fig:distri1} and \ref{fig:distri2} 
will be less patchy, and the posterior distributions can be 
constrained more precisely using more data. In order to avoid 
incorrect and/or ambiguous results using conventional frequentist 
fitting methods and tests, Bayesian inference could be and should 
be used in the observational and statistical studies of other 
compact objects and in general every fields of high-energy 
astrophysics, e.g., RL vs. millisecond pulsars, and 
Redback vs. Black Widow pulsars, etc.

\acknowledgments

We thank K.~S.~Cheng for valuable theoretical discussions. 
HFY thanks J.~Michael Burgess, Felix Ryde, and Yan-Ting Lam
for insightful discussions on Bayesian statistics. HFY 
acknowledges support from the Swedish National Space Board 
and the Swedish Research Council (Vetenskapsr\r{a}det). 
HFY is supported by the G\"{o}ran Gustafsson Foundation for 
Research in Natural Sciences and Medicine. CYH is supported 
by the National Research Foundation of Korea through grant 
2016R1A5A1013277. AKHK is supported by the Ministry of 
Science and Technology of the Republic of China (Taiwan) 
through grants 105-2112-M-007-033-MY2 and
105-2119-M-007-028-MY3. JT is supported by NSFC grants of 
Chinese Government under 11573010, U1631103 and 11661161010.

\vspace{5mm}

\facilities{\textit{Fermi} Gamma-ray Space Telescope/Large Area Telescope \citep{Atwood09}}
\software{{\tt PyMC3} \citep{Salvatier16}}

\end{document}